\documentclass[lettersize,journal]{IEEEtran}
\usepackage{cite}
\usepackage{amsmath,amssymb,amsfonts}
\usepackage{algorithmic}
\usepackage{graphicx}
\usepackage{textcomp}
\def\BibTeX{{\rm B\kern-.05em{\sc i\kern-.025em b}\kern-.08em
    T\kern-.1667em\lower.7ex\hbox{E}\kern-.125emX}}
\usepackage{booktabs}
\usepackage{xcolor}
\usepackage{epsfig}
\usepackage{epstopdf}
\usepackage[nolist]{acronym}
\graphicspath{{fig/}}
\usepackage{mathtools}
\usepackage{siunitx}
\usepackage{algorithmic}
\usepackage{algorithm}
\usepackage{tikzit}
\usepackage{enumitem}
\usepackage{subfig}
\usepackage{hyperref}

\tikzstyle{new style 0}=[fill=white, draw=black, shape=rectangle]
\tikzstyle{new style 1}=[fill={rgb,255: red,186; green,255; blue,174}, draw=black, shape=circle]

\tikzstyle{Arrow}=[->]
\tikzstyle{Dashed line}=[-, dashed, fill={rgb,255: red,243; green,250; blue,255}, opacity=0.8]
\tikzstyle{Algo}=[-, fill={rgb,255: red,244; green,255; blue,117}, draw=black]
\tikzstyle{Layer}=[-, fill={rgb,255: red,244; green,255; blue,117}]
\tikzstyle{new edge style 0}=[-]
\tikzstyle{Dashed Arrowed}=[<->, dashed]
\tikzstyle{Matrix}=[-, fill={rgb,255: red,206; green,255; blue,218}]
\tikzstyle{Vector}=[-, fill={rgb,255: red,226; green,225; blue,255}]
\tikzstyle{Bias}=[-, fill={rgb,255: red,255; green,245; blue,219}]

\usepackage{amsthm}

		\definecolor{applegreen}{rgb}{0.55, 0.71, 0.0}
\definecolor{awesome}{rgb}{1.0, 0.13, 0.32}
\definecolor{azure(colorwheel)}{rgb}{0.0, 0.5, 1.0}
\definecolor{darklavender}{rgb}{0.45, 0.31, 0.59}
\definecolor{cyan(process)}{rgb}{0.0, 0.72, 0.92}
\definecolor{brightmaroon}{rgb}{0.76, 0.13, 0.28}
\definecolor{ao(english)}{rgb}{0.0, 0.5, 0.0}
\definecolor{brightturquoise}{rgb}{0.03, 0.91, 0.87}
\definecolor{bondiblue}{rgb}{0.0, 0.58, 0.71}

\begin{acronym}
	\acro{ACDD}{Alamouti with cyclic delay diversity}
	\acro{URLLC}{ultra-reliable low-latency communications}
	\acro{3GPP}{third generation partnership project}
	\acro{PHY}{physical layer}
	\acro{MIMO}{multiple-input multiple-output}
	\acro{SIMO}{single-input multiple-output}
	\acro{MISO}{multiple-input single-output}
	\acro{SISO}{single-input single-output}
	\acro{MRC}{maximum-ratio combining}
	\acro{SNR}{signal-to-noise ratio}
	\acro{CP}{cyclic prefix}
	\acro{CDD}{cyclic delay diversity}
	\acro{FSC}{frequency-selective channel}
	\acro{STC}{space-time coding}
	\acro{FFT}{fast Fourier transform}
	\acro{LMMSE}{linear minimum mean-squared error}
	\acro{FER}{frame error rate}
	\acro{OFDM}{orthogonal frequency division multiplexing}
	\acro{OCDM}{orthogonal chirp division multiplexing}
	\acro{FSC}{frequency-selective channel}
	\acro{CSI}{channel state information}
	\acro{LMMSE-PIC}{linear minimum mean squared error with parallel interference cancellation}
	\acro{PFE}{perfect-feedback equalizer}
	\acro{FD}{frequency domain}
	\acro{PDP}{power delay profile}
	\acro{PDF}{probability density function}
	\acro{DFT}{discrete Fourier transform}
	\acro{ICI}{inter-carrier interference}
	\acro{OTFS}{orthogonal time frequency space}
	\acro{AWGN}{additive white Gaussian noise}
	\acro{SWH}{sparse Walsh-Hadamard}
	\acro{LLR}{log-likelihood ratio}
	\acro{PMF}{probability mass function}
	\acro{CRC}{cyclic redundancy check}
	\acro{PAM}{pulse amplitude modulation}
	\acro{QAM}{quadrature amplitude modulation}
	\acro{FWHT}{fast Walsh-Hadamard transform}
	\acro{MAP}{maximum a-posteriori}
	\acro{SC}{single-carrier}
	\acro{ISI}{inter-symbol interference}
	\acro{ZP}{zero-padding}
	\acro{BCJR}{Bahl, Cocke, Jelinek, and Raviv}
	\acro{ISAC}{integrated sensing and communication}
	\acro{DFRC}{dual function radar communication}
	\acro{UE}{user equipment}
	\acro{PSD}{power spectral density}
	\acro{BS}{base station}
	\acro{HPA}{high power amplifier}
	\acro{MIMO}{multiple-input, multiple-output}
	\acro{PAPR}{peak-to-average power ratio}
	\acro{AWGN}{additive white Gaussian noise}
	\acro{KPI}{key performance indicator}
	\acro{KPIs}{key performance indicators}  
	\acro{D2D}{device-to-device}
	\acro{ISAC}{integrated sensing and communication}
	\acro{DFRC}{dual-functional radar and communication}
	\acro{AoA}{angle of arrival}
	\acro{ToA}{time of arrival}
	\acro{EVM}{error vector magnitude}
	\acro{CPI}{coherent processing interval}
	\acro{eMBB}{enhanced mobile broadband}
	\acro{URLLC}{ultra-reliable low latency communications}
	\acro{mMTC}{massive machine type communications}
	\acro{QCQP}{quadratic constrained quadratic programming}
	\acro{BnB}{branch and bound}
	\acro{SER}{symbol error rate}
	\acro{LFM}{linear frequency modulation}
	\acro{ADMM}{alternating direction method of multipliers}
	\acro{BS}{base station}
	\acro{LoS}{line-of-sight}
	\acro{CFO}{carrier frequency offset}
	\acro{UL}{uplink}
	\acro{DL}{downlink}
	\acro{ULA}{uniform linear antenna array}
	\acro{MOO}{multi-objective optimization}
	\acro{IBO}{input back-off}
	\acro{QPSK}{quadrature phase shift keying}
	\acro{CCDF}{complementary cumulative distribution function}
	\acro{OFDM}{orthogonal frequency-division multiplexing }
	\acro{IoT}{internet of things}
	\acro{CRB}{Cram\'er-Rao bound}
	\acro{OFDM}{orthogonal frequency-division multiplexing}
	\acro{DL}{downlink}
	\acro{i.i.d}{independently and identically distributed}
	\acro{UL}{uplink}
	\acro{LoS}{line of sight}
	\acro{DFT}{discrete Fourier transform}
	\acro{HPA}{high power amplifier}
	\acro{IBO}{input power back-off}
	\acro{MIMO}{multiple-input, multiple-output}
	\acro{PAPR}{peak-to-average power ratio}
	\acro{AWGN}{additive white Gaussian noise}
	\acro{KPI}{key performance indicator}
	\acro{KPIs}{key performance indicators}  
	\acro{D2D}{device-to-device}
	\acro{RSU}{roadside unit}
	\acro{ISAC}{integrated sensing and communication}
	\acro{DFRC}{dual-functional radar and communication}
	\acro{AoA}{angle of arrival}
	\acro{AoAs}{angles of arrival}
	\acro{AoD}{angle of departure}
	\acro{AoDs}{angles of departure}
	\acro{ToA}{time of arrival}
	\acro{ToAs}{times of arrival}
	\acro{EVM}{error vector magnitude}
	\acro{CPI}{coherent processing interval}
	\acro{eMBB}{enhanced mobile broadband}
	\acro{URLLC}{ultra-reliable low latency communications}
	\acro{mMTC}{massive machine type communications}
	\acro{QCQP}{quadratic constrained quadratic programming}
	\acro{BnB}{branch and bound}
	\acro{SER}{symbol error rate}
	\acro{LFM}{linear frequency modulation}
	\acro{ADMM}{alternating direction method of multipliers}
	\acro{CCDF}{complementary cumulative distribution function}
	\acro{MISO}{multiple-input, single-output}
	\acro{CSI}{channel state information}
	\acro{OTFS}{orthogonal time frequency space}
	\acro{LDPC}{low-density parity-check}
	\acro{BCC}{binary convolutional coding}
	\acro{IEEE}{Institute of Electrical and Electronics Engineers}
	\acro{ULA}{uniform linear antenna}
	\acro{B5G}{beyond 5G}
	\acro{MOOP}{multi-objective optimization problem}
	\acro{MLD}{maximum likelihood}
	\acro{ML}{machine learning}
	\acro{FML}{fused maximum likelihood}
	\acro{RF}{radio frequency}
	\acro{AM/AM}{amplitude modulation/amplitude modulation}
	\acro{AM/PM}{amplitude modulation/phase modulation}
	\acro{BS}{base station}
	\acro{MM}{majorization-minimization}
	\acro{LNCA}{$\ell$-norm cyclic algorithm}
	\acro{OCDM}{orthogonal chirp-division multiplexing}
	\acro{TR}{tone reservation}
	\acro{COCS}{consecutive ordered cyclic shifts}
	\acro{LS}{least squares}
	\acro{CVE}{coefficient of variation of envelopes}
	\acro{BSUM}{block successive upper-bound minimization}
	\acro{ICE}{iterative convex enhancement}
	\acro{SOA}{sequential optimization algorithm}
	\acro{BCD}{block coordinate descent}
	\acro{SINR}{signal to interference plus noise ratio}
	\acro{MICF}{modified iterative clipping and filtering}
	\acro{PSL}{peak side-lobe level}
	\acro{SDR}{semi-definite relaxation}
	\acro{KL}{Kullback-Leibler}
	\acro{ADSRP}{alternating direction sequential relaxation programming}
	\acro{MUSIC}{multiple signal classification}
	\acro{EVD}{eigenvalue decomposition}
	\acro{SVD}{singular value decomposition}
	\acro{MSE}{mean squared error}
	\acro{SNR}{signal-to-noise ratio}
	\acro{CRB}{Cram\'er-Rao bound}
	\acro{QPSK}{quadrature phase shift keying}
	\acro{RCS}{radar cross-section}
	\acro{NOMA}{non-orthogonal multiple access}
	\acro{IRS}{intelligent reflecting surface}
	\acro{IoT}{Internet of things}
	\acro{UAV}{unmanned aerial vehicle}
	\acro{V2X}{vehicle-to-everything}
	\acro{SDRs}{software-defined radios}
	\acro{FDD}{frequency-division duplex}
	\acro{NR}{new radio}
	\acro{FIM}{Fisher information matrix}
	\acro{PDF}{probability density function}    
	\acro{CRLB}{Cram\'er-Rao Lower Bound}
	\acro{CORDIC}{coordinate rotation digital computer}
	\acro{CIR}{channel impulse response}
	\acro{TO}{timing offset}
	\acro{FO}{frequency offset}
	\acro{CPI}{coherent processing interval}
	\acro{FFT}{fast Fourier transform}
	\acro{IFFT}{inverse fast Fourier transform}
	\acro{DFT}{discrete Fourier transform}
	\acro{DoA}{direction of arrival}
	\acro{MLP}{multi-layer perceptron}
	\acro{NN}{neural network}
	\acro{DL}{deep learning}
	\acro{MLE}{maximum likelihood estimator}
	\acro{TDD}{time-division duplexing}
	\acro{LWA}{leaky-wave antenna}
	\acro{mmWave}{milli-meter wave}
\end{acronym}

\DeclareMathOperator{\SNR}{SNR}

\DeclareMathOperator{\dB}{dB}

\DeclareMathOperator{\res}{res}

\DeclareMathOperator{\rad}{rad}
\DeclareMathOperator{\diag}{diag}
\DeclareMathOperator{\CRB}{CRB}
\DeclareMathOperator{\vectorize}{vec}
\DeclareMathOperator{\TSVD}{TSVD}

\DeclareMathOperator{\otherwise}{otherwise}
\DeclareMathOperator{\subjectto}{subject \ to}

\DeclareMathOperator*{\argmin}{arg\,min}

\usepackage{soul,color}

\usepackage{xcolor,cite,etoolbox}
\makeatletter
\pretocmd\@bibitem{\color{black}\csname keycolor#1\endcsname}{}{\fail}
\newcommand\citecolor[1]{\@namedef{keycolor#1}{\color{black}}}
\makeatother

\citecolor{6587285}
\citecolor{8123841}
\citecolor{10279367}
\citecolor{10250200}
\citecolor{5336796}
\citecolor{7123663}
\citecolor{8292713}
\citecolor{9664489}
\citecolor{pmlr-v97-rahaman19a}
\citecolor{9540344}
\citecolor{10038611}
\citecolor{9828505}
\citecolor{10061453}
\citecolor{8386661}
\usepackage{fancyhdr}
\fancypagestyle{firststyle}{
	\fancyhf{}
	\fancyhead[L]{S. Naoumi, A. Bazzi, R. Bomfin, and M. Chafii, ``Complex Neural Network based Joint AoA and AoD Estimation for Bistatic ISAC'', accepted in \textit{IEEE Journal of Selected Topics in Signal Processing}, March. 2024.}

}
\begin{document}
%
%
% \title{Complex Neural Network based Joint AoA and AoD Estimation for Bistatic ISAC}
%
\title{
Complex Neural Network based Joint AoA and AoD Estimation for Bistatic ISAC
\thanks{
Salmane Naoumi is with NYU Tandon School of Engineering, Brooklyn, 11201, NY (e-mail: sn3397@nyu.edu).

Roberto Bomfin is with the Engineering Division, New York University (NYU) Abu Dhabi, UAE (e-mail: rcd9059@nyu.edu ). 

Ahmad Bazzi and Marwa Chafii are with the Engineering Division, New York University (NYU) Abu Dhabi, UAE and NYU WIRELESS, NYU Tandon School of Engineering, Brooklyn, NY (e-mail: ahmad.bazzi@nyu.edu, marwa.chafii@nyu.edu).
}
}
\author{Salmane Naoumi, Ahmad Bazzi, Roberto Bomfin, Marwa Chafii}        

\maketitle
% The paper headers
\markboth{submitted, 2023}%
{Shell \MakeLowercase{\textit{et al.}}: A Sample Article Using IEEEtran.cls for IEEE Journals}

\IEEEpubid{}
\thispagestyle{firststyle}

\begin{abstract}
Integrated sensing and communication (ISAC) in wireless systems has emerged as a promising paradigm, offering the potential for improved performance, efficient resource utilization, and mutually beneficial interactions between radar sensing and wireless communications, thereby shaping the future of wireless technologies. In this work, we present two novel methods to address the joint angle of arrival and angle of departure estimation problem for bistatic ISAC systems. Our proposed methods consist of a deep learning (DL) solution leveraging complex neural networks, in addition to a parameterized algorithm. By exploiting the estimated channel matrix and incorporating a preprocessing step consisting of a coarse timing estimation, we are able to notably reduce the input size and improve the computational efficiency. In our findings, we emphasize the remarkable potential of our DL-based approach, which demonstrates comparable performance to the parameterized method that explicitly exploits the multiple-input multiple-output (MIMO) model, while exhibiting significantly lower computational complexity.
\end{abstract}

\section{Introduction}\label{sec:introduction}
The advent of \ac{ISAC} marks a transformative breakthrough for the realm of 6G networks \cite{chafii2023twelve}. While radar sensing and wireless communications have traditionally progressed independently, the emergence of joint communication and sensing now brings forth a game-changing paradigm, presenting unprecedented opportunities to revolutionize spectrum efficiency, reduce hardware cost and power consumption, and redefine how networks perceive and interact with their surroundings. 
Given this context, \ac{ISAC} has attracted significant research interest and attention from both academic and industrial sectors. Indeed, various scenarios and challenges have been explored. These include, \ac{DFRC} \cite{LiuTSP}, unmanned aerial vehicles \cite{akhil17}, waveform design \cite{akhil16}, beamforming design \cite{akhil18}, security \cite{10373185,10437283}, and intelligent reflecting surfaces \cite{akhil15, dexinRIS}.

Communication-centric \ac{ISAC}, which is based on the utilization of an optimized communication waveform for sensing tasks, is a promising research area that offers the potential for fruitful outcomes by leveraging transmitted communication signals to enhance sensing capabilities. Repurposing the estimation process of channel state information provides an opportunity to accurately determine physical sensing parameters like the \ac{DoA}, range, and speed of targets of interest. Indeed, numerous research studies have examined and validated the utilization of bistatic radar topology in conjunction with \ac{OFDM}-based signal processing to address communication-aided sensing optimization \cite{introCopp,akhil28, akhil37}. Notably, they have explored this approach within a practical bistatic setup \cite{introOFDM}, demonstrating its effectiveness in accurately detecting and estimating target parameters such as delay and Doppler shift. 
Furthermore, a power budget analysis was conducted in \cite{RobertoPower} to study the behavior of communication and radar \acp{SNR} as function of distance.
Nevertheless, processing the received communication signal to achieve high sensing quality remains practically challenging, as both communications and radar systems are not mutually optimized for the sensing function.
Various challenges arise when designing the radar signal processing functions, including high peak-to-average power ratio \cite{akhil22}, clutter noise, multi-path reflections, unoptimized side-lobes, and Doppler ambiguity caused by the \ac{CP} \cite{IntroWaveform}. 

In this work, we investigate parameter estimation in a bistatic radar setup within an \ac{ISAC} system, where we focus on estimating both \ac{AoA} and \ac{AoD}. Since the maximum likelihood estimator is a highly multidimensional optimization problem, which is infeasible to be realized in real-time with practical hardware resources, we propose to leverage \acp{NN} ability of being universal approximators, to perform this complex estimation problem with low complex optimized \ac{NN} architectures.
In the literature \cite{surveySa}, the \ac{AoA} and \ac{AoD} estimation problem is either framed as a regression problem \cite{survey17, survey20} by training architectures like the \ac{MLP} to minimize an objective function such as the \ac{MSE} between estimates and true angles, or as a spectrum based sensing estimation \cite{survey29, survey33} where an \ac{NN} is trained on hot encoded ground-truth vectors representing targets likelihoods at given angles of a discrete grid with a given cardinality.
In the latter case, the network does not rely on prior knowledge of the number of targets. Instead, it is trained to output an estimated spectrum and subsequently requires an additional step for target detection to extract angle estimates from peaks, as commonly employed in conventional methodologies like MUSIC. Other \ac{DL} approaches have been proposed such as a multi-stage DL \cite{survey18} consisting of a multitask autoencoder and a series of parallel multi-layer classifiers. In our study, we present a novel method addressing the sensing estimation problem that capitalizes on a distinct input and trains complex-valued \acp{NN} to minimize the \ac{MSE} objective function on training datasets generated using the ideal system model. 
Moreover, we propose a parameterized algorithm that exploits the knowledge of the model to be used for benchmarking. More specifically, the parameterized $2$D method estimates the \ac{AoA} and \ac{AoD} of the targets utilizing full knowledge of array steering vectors and \ac{OFDM} sub-carrier regular structure. 
\textcolor{black}{Although \ac{OFDM} is assumed in the model, our work remains valid for any precoded \ac{OFDM} modulation such as \ac{DFT}-spread OFDM.}
For comparison, the \ac{CRB} is also shown.
The results reveal that the DL-based solution provides a comparable estimation performance to the parameterized method, while requiring considerably less computational efforts.

The main contributions of this work are summarized as follows.
\begin{itemize}
	\item \textbf{Novel joint \ac{AoA}/\ac{AoD} \ac{DL} estimator.} We present a new \ac{DL} technique for sensing parameter estimation, i.e. the \acp{AoA} and \acp{AoD} in an \ac{ISAC} system. 
	Departing from conventional input sources, the method utilizes the estimated channel matrix, which captures essential information regarding the sensing parameters. 
	The \ac{DL} model employs complex-valued neural networks, and includes complex convolutional and linear layers, with weights represented as complex matrices, enhancing the model's ability to accurately represent the channel matrix. The complex rectified linear unit (CReLU) is chosen as the activation function. 
	The network structure comprises three hidden layers and a final output layer predicting the sensing parameters. Moreover, the \ac{MSE} serves as the objective function during the training phase.
	
	\item \textbf{Novel design of a parameterized \ac{AoA}/\ac{AoD} estimator.} We develop a parameterized method for \ac{AoA}/\ac{AoD} estimator, which exploits the complete model knowledge, including the array geometries. 
	The method involves transforming the received channel matrix by leveraging the \ac{ULA} structures at transmit and receive sides into smaller channel matrices with given sub-array sizes.
	The method makes use of the induced Hankel structure to further compute certain eigenvalues, which aid in the \ac{AoD} estimation. 
	Regarding \ac{AoA} estimation, a \ac{LS} fit is required at a later stage of the algorithm.
\end{itemize}

Furthermore, we unveil some important insights, i.e.
\begin{itemize}
	\item 
	Through simulations, we study the impact of training \ac{SNR} on the achieved \ac{MSE} performance. 
	More specifically, the \ac{MLP} architecture is trained using simulation data generated at specific \acp{SNR}, and the network's performance is evaluated on a test dataset spanning a range of given \ac{SNR} values.
	We observe that loss curves steadily decrease during training, indicating effective learning and generalization to unseen data. 
	Regarding overfitting, we employ a learning rate schedule.
	In addition, we showcase various simulation results illustrating the superior performance and potential of the proposed methods compared, and their closeness to the \ac{CRB} benchmarks. 
	
    \item We conduct a comprehensive computational complexity analysis of the two proposed methods, focusing on the total number of multiplications and additions required for their implementation. This analysis includes a detailed examination of the complexities associated with the \ac{DL} technique, which employs architectures with three hidden layers, by counting the complexities involved in its different operations. Moreover, for the parameterized approach, we provide a breakdown of the computational complexities associated with specific blocks, such as channel estimation, coarse timing estimation, and sensing estimation. This includes highlighting the estimation overheads and detailing the total number of operations for each block. We also present the maximum likelihood criterion for the estimation problem at hand, emphasizing the significant complexity reductions achieved by both methods compared to the \ac{MLE}. Additionally, our simulations demonstrate that, in comparison to the parametric $2$D estimation procedure, the proposed \ac{DL} technique enjoys less computational complexity. 
For example, with $8$ receive antennas, the $2$D algorithm requires $6.5$ times more multiplications than the \ac{DL} one, which is $15.27$\% of the computational complexity required by the $2$D algorithm for multiplications.
Despite a performance degradation, especially at higher \ac{SNR} values, these results highlight the significant percentage increases in computational efficiency and scalability provided by the \ac{DL} design. 

    \item We investigate the influence of antenna radiation patterns on the performance of bi-static \ac{ISAC} systems. After incorporating these patterns into our system model, we explore how changes in their parameters impact the \ac{CRB}. Our analysis reveal that narrower, more focused beams require higher \acp{SNR} to attain a predetermined \ac{CRB} for target sensing applications. This insight emphasizes the complex interplay between antenna design choices and the performance of \ac{ISAC} systems, highlighting the importance of integrating the radiation pattern characteristics to enhance the sensing accuracy in realistic applications. 
    
\end{itemize}

% \textcolor{black}{\input{Actions/list-of-contributions.tex}}
% \textcolor{black}{\input{Actions/list-of-insights.tex}}
The detailed structure of the following paper is given as follows. 
In Section \ref{sec:system_model}, we introduce the communication-centric \ac{ISAC} system model, along with the channel estimator and coarse timing estimate used throughout the paper.
Section \ref{sec:MLE} derives the maximum likelihood estimator of the estimation problem.
 In Section \ref{sec:ML_method}, we present our machine learning-based AoA and AoD estimator, accompanied with its computational complexity.
Moreover, Section \ref{sec:benchmarking_algorithm} presents our $2$D parametrized estimator, along with its detailed computational complexity analysis.
 Section \ref{sec:simulations} provides numerical results to verify our analysis before concluding the paper in Section \ref{sec:conclusion}.
 
\textbf{Notation}: Upper-case and lower-case boldface letters denote matrices and vectors, resp. $(\cdot)^T$, $(\cdot)^*$, $(\cdot)^H$ and $(\cdot)^\dagger$ represent the transpose, the conjugate, the transpose-conjugate, and the pseudo-inverse operators. We denote by $*$ the convolution operator. 
For any complex number $z \in \mathbb{C}$, the magnitude is denoted as $\vert z \vert$, and its angle is $ \arg(z)$. The real part of $z$ is denoted as $\Re(z)$, whereas the imaginary part is denoted as $\Im(z)$. The $\ell_2$ norm of a vector $\pmb{x}$ is denoted as $\Vert \pmb{x} \Vert$. The matrices  $\pmb{F}$ and $\pmb{I}$ are the Fourier and the identity matrices with appropriate dimensions, resp. 
For matrix indexing, the $(i,j)^{th}$ entry of matrix $\pmb{A}$ is denoted by $[\pmb{A}]_{[i,j]}$ and its $j^{th}$ column is denoted as $\pmb{A}_{[:,j]}$. The operator $\otimes$ is the \textit{Kronecker} product.
The big-$\mathcal{O}$ notation is $\mathcal{O}()$.
\textcolor{black}{The $\res_{M,N}(\pmb{x})$ is the reshape operator which returns an array of  dimensions $M \times N$ with the same entries as the input data $\pmb{x}$.}\textcolor{black}{Also, the operator $\bmod_y(x)$ denotes the remainder of the division $x / y$.}
%For a set of integers $\mathcal{S}$, the set $\mathcal{S} + n = \lbrace s + n, \forall s \in \mathcal{S} \rbrace$. The vector $\pmb{e}_{N}$ is an $N \times 1$ vector given as $\pmb{e} = \begin{bmatrix}
%0 & 1 & \ldots & N-1  \end{bmatrix}^T$. The vector $\pmb{p}_i$ is a vectors of all-zeros except $1$ it it's $i^{th}$ entry.

    \section{System Model}\label{sec:system_model}
\begin{figure}[!t]
\centering
\includegraphics[width=3.5in]{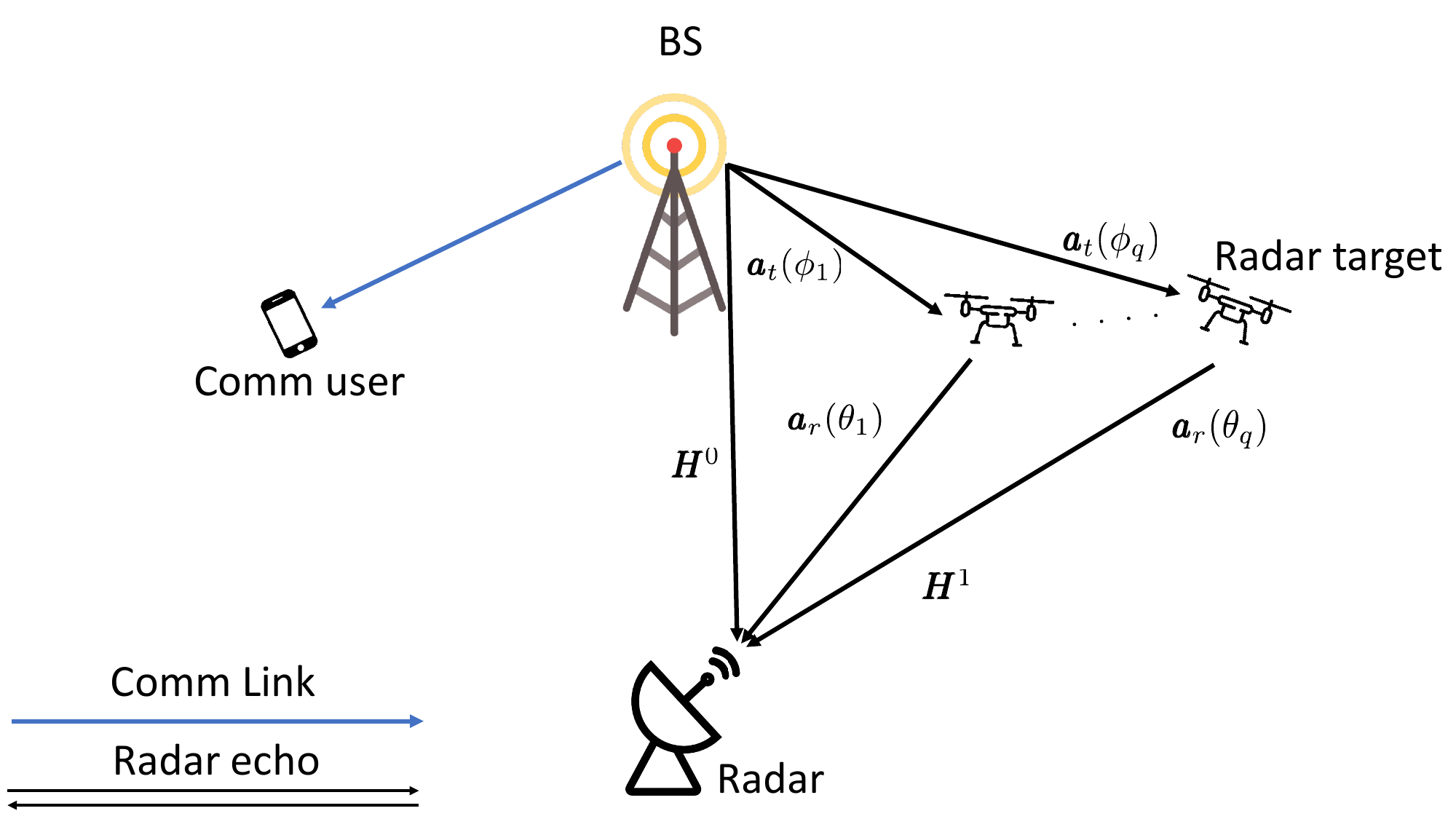}
\caption{An \ac{ISAC} scenario composed of $q$ targets in a bistatic fashion.}
\label{fig_1}
\end{figure}
Consider a \ac{BS} with $N_t$ antennas transmitting \ac{OFDM} symbols in the downlink. The \ac{OFDM} symbol duration, excluding the \ac{CP}, is denoted as $T = \frac{1}{\Delta_f}$, where $\Delta_f$ is the sub-carrier spacing. Before expressing the transmit symbols, we describe the wireless channel related to the bistatic radar.
\textcolor{black}{The \ac{ISAC} framework can be classified into three  categories: \textit{communication-centric}, \textit{radar-centric}, and \textit{joint design} \cite{9540344}, \cite{10038611}. 
This paper primarily focuses on the communication-centric class of \ac{ISAC}, whereby radar sensing can be considered as an "add-on" to a communication system \cite{KonpalISAC}. 
The key objective of this design approach is to utilize communication waveforms for extracting radar information through signal processing of target echoes. 
On the other hand, in radar-centric design, the approach involves modulating information signaling within established radar waveforms, such as chirping \cite{9828505}. The joint design category encompasses systems that are collaboratively designed from the outset to achieve a flexible balance between sensing and communication performance \cite{10061453}, \cite{8386661}.
As this paper explores a scenario where a communication signal is transmitted to each user by a \ac{BS} primarily dedicated to communications, and the radar receiver utilizes the backscattered communication signal for precise estimation of target sensing parameters, the configuration of \ac{ISAC} can effectively be associated to the communication-centric category.
Furthermore, we note that we are using preamble knowledge based on communication standards to perform sensing tasks, hence an \ac{ISAC} integration. 
The literature contains some work relating to communication-centric \ac{ISAC}.
For example, a mono-static setting of this \ac{ISAC} setting has been described in \cite{10279367} for \ac{OTFS} digital modulation.
Moreover, a \ac{TDD} massive \ac{MIMO} system was adopted in \cite{10250200}, where a \ac{BS} serving communication users, while profiting from the scattered signal to sense the environment.}To this extent, the system model is represented in Fig.~\ref{fig_1}.
\subsection{Bistatic Radar Channel Model}
\label{sec:bistatic-radar-channel-model}
Under the simplifying assumptions of no clutter, no frequency and time offsets between the BS and the radar unit, and fixed targets, the general \ac{CIR} between the $n_t^{th}$ transmit antenna at the \ac{BS} and the $n_r^{th}$ receive antenna at the radar unit can be expressed as 
\begin{equation}
\label{eq:CIR}
\begin{split}
\textcolor{black}{h_{n_r,n_t}(t,\tau) 
=
\sum\limits_{m=0}^{q}
\alpha_m
\bar{a}_{n_r}(\theta_m)
\bar{a}_{n_t}(\phi_m) \delta(\tau - \tau_m ),}
\end{split}
\end{equation} 
where $n_r = 1 \ldots N_r$ and $n_t = 1 \ldots N_t$ for $N_r$ receive antennas, and $N_t$ transmit antennas. In addition, $\alpha_m$ and $ \tau_m$ denote the signal attenuation coefficient and delay of the $m^{th}$ reflection, respectively, and are assumed to be time-invariant during the parameter estimation period. %Note that the delay is the round-trip delay between the \ac{BS} and target in addition to the delay between target and radar. 
The function $\delta(.)$ is the Dirac delta function, and $q$ is the number of targets. Furthermore, $\theta_m$ represents the \ac{AoA} between the $m^{th}$ target and the radar unit, and $\phi_m$ is the \ac{AoD} between the \ac{BS} and the $m^{th}$ target.
\textcolor{black}{Moreover, the quantity $\bar{a}_{n_t}(\phi)$ is the \textit{actual steering coefficient} of the $n_t^{th}$ transmit antenna at the \ac{BS} towards an angle $\phi$.
Likewise, $\bar{a}_{n_r}(\theta)$ is the actual steering coefficient of the $n_r^{th}$ receive antenna at the radar towards an angle $\theta$.}
\textcolor{black}{Typically, antennas have the capability to emit energy in specific directions, allowing for targeted radiation within a defined region of space. This is in contrast to omni-directional antennas, which evenly perceive signals from all spatial coordinates.
To quantify this, we resort to the antenna radiation pattern, which is defined as a complex function of direction, whose value gives the intensity of the radiated field in the far field area \cite{6587285}.
Assuming identical, but unknown, transmit antennas are used at the \ac{BS}, the transmit steering vector can be defined as \cite{8123841}
\begin{equation}
\label{eq:actual-steering-Tx-vector}
	\bar{a}_{n_t}(\phi)
	=
	g_t(\phi)
	a_{n_t}(\phi),
\end{equation}
where $g_t(\phi)$ is the transmit antenna radiation pattern at angle $\phi$. Similarly, the receive steering vector can be also defined as 
\begin{equation}
\label{eq:actual-steering-Rx-vector}
	\bar{a}_{n_r}(\theta)
	=
	g_r(\theta)
	a_{n_r}(\theta),
\end{equation}
where $g_r(\theta)$ is the receive antenna radiation pattern, also assumed to be unknown.
Note that in the expressions found in equations \eqref{eq:actual-steering-Tx-vector} and \eqref{eq:actual-steering-Rx-vector}, the actual transmit and receive steering vectors are written as a function of the hypothetical isotropic steering vectors, namely $a_{n_t}(\phi)$ and $a_{n_r}(\theta)$.
Said differently, setting $g_t(\phi)$ (similarly $g_r(\theta)$) to unity, we coincide with the hypothetical isotropic case.}
Note that the first path, i.e. $m=0$, represents the \ac{LoS} path between the \ac{BS} and the radar unit.
The channel given in \eqref{eq:CIR} can be expressed in the discrete frequency domain on the $n^{th}$ sub-carrier, in matrix form, as
\begin{equation}
\label{eq:CIR-freq}
\textcolor{black}{\pmb{H}_n
=
\bar{\pmb{A}_r}(\pmb{\Theta})
\pmb{\Psi}
\pmb{D}_n(\pmb{\tau})
\bar{\pmb{A}_t}^T(\pmb{\Phi}) \in \mathbb{C}^{N_r \times N_t},}
\end{equation}
\textcolor{black}{where
\begin{equation}
\label{eq:bar_A_t}
	\bar{\pmb{A}_t}(\pmb{\Phi}) 
	= 
	\begin{bmatrix} 
	\bar{\pmb{a}}_t(\phi_0) & \bar{\pmb{a}}_t(\phi_1) &  \ldots & \bar{\pmb{a}}_t(\phi_q) 
	\end{bmatrix},
\end{equation}
and
\begin{equation}
\label{eq:bar_A_r}
	\bar{\pmb{A}_r}(\pmb{\Theta}) 
	= 
	\begin{bmatrix} 
	\bar{\pmb{a}}_r(\theta_0) & \bar{\pmb{a}}_r(\theta_1) & \ldots & \bar{\pmb{a}}_r(\theta_q) 
	\end{bmatrix},
\end{equation}
are the \textit{actual steering matrices} resulting from the \acp{AoD} between the \ac{BS}-targets and the \acp{AoA} between targets-radar, respectively.
Moreover, $\pmb{\Psi} =\diag ( 	\begin{bmatrix} \alpha_0  & \ldots & \alpha_q \end{bmatrix} )$ includes the signal attenuation coefficients.
}
Also, $	\pmb{D}_n(\pmb{\tau}) =\diag (\begin{bmatrix} c_n(\tau_0)  & \ldots & c_n(\tau_q) \end{bmatrix})$ is a matrix containing all delays $c_n(\tau) = e^{-j 2 \pi n \Delta_f \tau}$. In the next sub-section, we describe the received signal in the frequency domain as seen by the radar.
\textcolor{black}{Using equations \eqref{eq:actual-steering-Tx-vector}, \eqref{eq:actual-steering-Rx-vector}, we can write the actual steering matrices in terms of their hypothetical counterparts as 
\begin{equation}
	\label{eq:bar_A_t_2}
	\bar{\pmb{A}_t}(\pmb{\Phi}) 
	=
	\pmb{G}_t(\pmb{\Phi})
	\pmb{A}_t(\pmb{\Phi}),
\end{equation}
and
\begin{equation}
	\label{eq:bar_A_r_2}
	\bar{\pmb{A}_r}(\pmb{\Theta}) 
	=
	\pmb{G}_r(\pmb{\Theta})
	\pmb{A}_r(\pmb{\Theta}),
\end{equation}
where 
$\pmb{G}_t(\pmb{\Phi}) = \diag \big( \begin{bmatrix} g_t(\phi_0) & \ldots & g_t(\phi_q) \end{bmatrix} \big)$ 
and 
$\pmb{G}_r(\pmb{\Theta}) = \diag \big( \begin{bmatrix} g_r(\theta_0) & \ldots & g_r(\theta_q) \end{bmatrix} \big)$ are matrices arising due to the impact of the antenna radiation pattern at the arrays placed at the \ac{BS} and the radar, respectively. 
Moreover, $\pmb{A}_t(\pmb{\Phi})$ and $\pmb{A}_r(\pmb{\Theta})$ are the hypothetical steering matrices, which are defined similar to $\bar{\pmb{A}_t}(\pmb{\Phi})$ and $\bar{\pmb{A}_r}(\pmb{\Theta})$ found in equations \eqref{eq:bar_A_t} and \eqref{eq:bar_A_r}, respectively.
Invoking \eqref{eq:bar_A_t_2} and \eqref{eq:bar_A_r_2} in \eqref{eq:CIR-freq} and applying the commutativity property of diagonal matrices (i.e. $\pmb{YZ} = \pmb{ZY}$ for any two diagonal matrices $\pmb{Y}$ and $\pmb{Z}$), we can encapsulate the model as follows
\begin{equation}
	\pmb{H}_n
	=
	\pmb{A}_r(\pmb{\Theta})
	\pmb{G}
	\pmb{D}_n(\pmb{\tau})
	\pmb{A}_t^T(\pmb{\Phi}),
\end{equation}
where $\pmb{G}$ jointly contains attenuation components, along with transmit and receive antenna radiation patterns, i.e. $\pmb{G} =  \pmb{G}_r(\pmb{\Theta})\pmb{\Psi} \pmb{G}_t(\pmb{\Phi})$. 
Note that $\pmb{G}$ preserves its diagonal structure, and can be expressed as
$\pmb{G} = \diag \big( \begin{bmatrix}
	\alpha_0g_t(\phi_0)g_r(\theta_0) &
	\ldots &
	\alpha_qg_t(\phi_q)g_r(\theta_q)
\end{bmatrix} \big)$.
For simplicity, we denote $\breve{\alpha}_k = \alpha_kg_t(\phi_k)g_r(\theta_k)$ and treat $\breve{\alpha}_k$ as an unknown quantity incorporating the channel unknown components, and the transmit/receive antenna radiation patterns.
}
\subsection{Radar Signal Model}
The \ac{BS} transmits $K$ \ac{OFDM} signals serving communication users in the scene. The $k^{th}$ \ac{OFDM} can be expressed as
\begin{equation}
	\label{eq:tx-OFDM-DL}
	\pmb{x}_{k}(t) = 
	\sum\limits_{n = 1 }^N
	\pmb{s}_{n,k} 
        c_n(-t)
	\Pi(t- k T_o), \quad \forall k = 1 \ldots K,
\end{equation}
where $N$ is the number of active sub-carriers occupying each of the $K$ \ac{OFDM} symbols and $T_o = T + T_{CP}$ is the over-all \ac{OFDM} symbol duration. Moreover, $\Pi(t)$ is the windowing function. We assume an ideal rectangular function, that is
\begin{equation}
\label{eq:ideal-window}
	\Pi(t)
	=
	\begin{cases}
		1, & t \in [-T_{\text{CP}},T] \\
		0, & \otherwise,
	\end{cases}
\end{equation}
where $T_{\text{CP}}$ is the \ac{CP} duration, which should be greater than the maximum of propagation delays in order to guarantee a cyclic convolution with the channel.
Furthermore, $\pmb{s}_{n,k} \in \mathbb{C}^{N_t \times 1}$ is the modulated symbol onto the $n^{th}$ \ac{OFDM} sub-carrier within the $k^{th}$ \ac{OFDM} symbol.
\textcolor{black}{We note that in case of precoded OFDM, $\pmb{s}_{n,k}$ simply represents the resulting signal to be transmitted in the corresponding $n^{th}$ sub-carrier after precoding the data symbols associated with the $k^{th}$ transmission block.}
Combining \eqref{eq:CIR-freq}, \eqref{eq:tx-OFDM-DL}, \eqref{eq:ideal-window} and applying \ac{FFT}, the radar unit reads the following data on the $n^{th}$ subcarrier and $k^{th}$ \ac{OFDM} symbol,
\begin{equation}
    \label{eq:system-model-obs}
	\pmb{y}_{n,k}
	=
	\pmb{H}_n
	\pmb{s}_{n,k}
	+
	\pmb{w}_{n,k}
	\in 
	\mathbb{C}^{N_r \times 1}.
\end{equation}
\textcolor{black}{The vector $\pmb{w}_{n,k}$ is \ac{AWGN} on the $n^{th}$ sub-carrier and $k^{th}$ symbol, with zero mean and covariance $\sigma^2 \pmb{I}$.}In this paper, we focus on a communication-centric \ac{ISAC} problem, where an existing infrastructure is performing communication tasks, whereas a radar unit is installed to estimate the sensing parameters of the different targets.

\subsection{Channel Estimation}
\label{subsec:channel-estimation}
In this sub-section, we describe the channel estimates that are used for both \ac{DL}-based and parameterized methods to estimate \ac{AoA} and \ac{AoD}. 
Assuming $K_P$ OFDM symbols occupying $N_P$ sub-carriers each are being transmitted, and assuming that the radar unit has knowledge about the data symbols being transmitted, the channel estimation is performed according to the well-known least squares as follows 
\begin{equation}
	\label{eq:method2_step0}
	\bar{\pmb{H}}_n = \pmb{Y}_{n} \pmb{S}_{P,n}^H (\pmb{S}_{P,n}\pmb{S}_{P,n}^H)^{-1},
\end{equation}
where $\pmb{Y}_n$ contains all pilot \ac{OFDM} symbols, i.e. $\lbrace   \pmb{y}_{n,k}  \rbrace_{k=1}^{K_P}$. Likewise, $\pmb{S}_{P,n}$ contains the known pilot information, $\lbrace \pmb{s}_{n,k} \rbrace_{k=1}^{K_P}$. Note that $\bar{\pmb{H}}_n = \pmb{H}_n + \pmb{W}_{P,n} \pmb{S}_{P,n}^H (\pmb{S}_{P,n}\pmb{S}_{P,n}^H)^{-1}$. It is more convenient to express \ac{CSI} estimates as follows
\begin{equation}
\begin{split}
        \label{eq:method2_step0_1}
	\pmb{\bar{{H}}}
	=
	\begin{bmatrix}
		\vectorize(\bar{\pmb{H}}_1) & 
		\vectorize(\bar{\pmb{H}}_2) &
		\ldots &
		\vectorize(\bar{\pmb{H}}_{N_P})
	\end{bmatrix},
\end{split}
\end{equation}
where $\pmb{\bar{{H}}} \in \mathbb{C}^{N_t N_r \times N_P}$ contains the frequency domain channel in its rows for all combinations of transmit/receive antenna pairs.
\textcolor{black}{We can express $\pmb{\bar{{H}}}$ in terms of its true and noise counterparts as 
\begin{equation}
	\pmb{\bar{{H}}}
	=
	\pmb{{{H}}}
	+
	\pmb{\bar{{W}}},
\end{equation}
where $\pmb{{{H}}}
	=
	\begin{bmatrix}
		\vectorize({\pmb{H}}_1) & 
		\ldots &
		\vectorize({\pmb{H}}_{N_P})
	\end{bmatrix}$ and 
\begin{equation}
	\label{eq:Wbar-definition}
	\pmb{\bar{{W}}}
	=
	\begin{bmatrix}
		\vectorize(\pmb{W}_{P,1} \pmb{S}_{P,1}^\dagger) & 
		\ldots &
		\vectorize(\pmb{W}_{P,N_P} \pmb{S}_{P,N_P}^\dagger)
	\end{bmatrix},
\end{equation}
where, for short, $\pmb{S}_{P,n}^\dagger$ stands for the pseudo-inverse of $\pmb{S}_{P,n}$, i.e. $\pmb{S}_{P,n}^\dagger=\pmb{S}_{P,n}^H (\pmb{S}_{P,n}\pmb{S}_{P,n}^H)^{-1}$ for all $n$.}

\subsection{Coarse Timing Estimation}
\label{subsec:coarse-timing-estimation}
The size of the channel estimation matrix in \eqref{eq:method2_step0_1} can be reduced in order to estimate \ac{AoA} and \ac{AoD}, such that the input size of the \ac{ML} algorithm is decreased.
Since each row of $\pmb{\bar{{H}}}$ in \eqref{eq:method2_step0_1} represents the frequency domain channel over a given transmit and receive antenna pair, we can take the IFFT of \eqref{eq:method2_step0_1} over its rows, and find the peaks corresponding to the time delay of a target reflection, which can be interpreted as a coarse timing estimation for the delay as a multiple of the sampling time. Notice that the \ac{AoA} and \ac{AoD} have a direct relation with the phase of the peaks in the time domain, thus, a matrix of size $N_t \times N_r$ with the peaks can be used to estimate these parameters instead of the larger matrix of \eqref{eq:method2_step0_1}. This method is briefly shown in the following.
The \acp{IFFT} per transmit/receive antenna pairs is performed as
% Fig.~\ref{fig_peaks_find} illustrates an example of channel impulse response estimation, where the peaks correspond to three different targets in the environment.

%
% \begin{figure}[!t]
% \centering
% \includegraphics[width=3.5in]{./figs/ch_IR_estimation2.png}
% \caption{Example of the channel impulse response for a setting with three targets at an \ac{SNR} = 30 dB for an arbitrary TX and RX antenna pair with $N_P=64$ sub-carriers. \hl{Use $\Re$ and $\Im$ in the legend to remain notation-consistent.}} 
% \label{fig_peaks_find}
% \end{figure}
\begin{figure}[t]
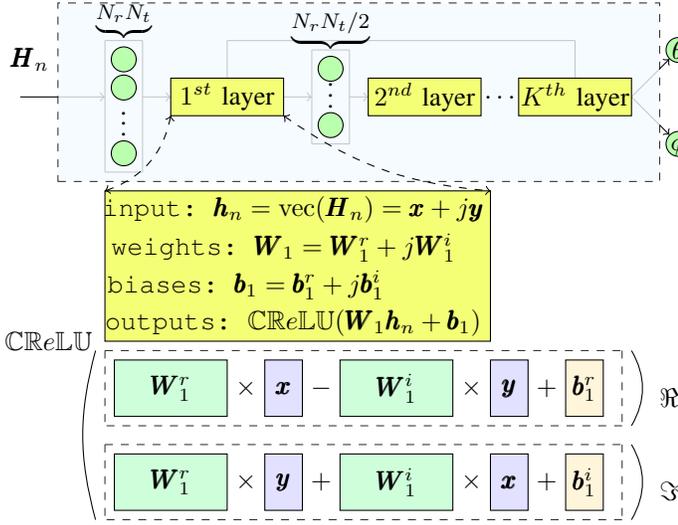

	\centering
	\ctikzfig{NN}
	\caption{Architecture of the proposed feed-forward \ac{NN} composed of complex-valued linear layers and $\mathbb{C R} e \mathbb{L U}$ activation functions. The channel matrix at each peak is used to compute the corresponding AoA and AoD.}
	\label{fig:NN}
\end{figure}

\begin{equation}
\label{eq:IFFT-step}
   \pmb{\bar{{h}}}_n = {\rm IFFT}( \pmb{\bar{{H}}}_{[n,:]} )
   = \pmb{F}^H \pmb{\bar{{H}}}_{[n,:]}^T,
\end{equation} 
$\forall n = 1 \ldots N_rN_t$. Then, the task is basically to find the indexes $k$ corresponding to the $q$ most likely peaks over all antenna pairs $|\pmb{\bar{{h}}}_{n[k]}|$ from $\pmb{\bar{{h}}}_n = [\pmb{\bar{{h}}}_{n{[1]}}\cdots \pmb{\bar{{h}}}_{n{[N_P]}}]\in \mathbb{C}^{N_P}$, which are defined as
% $\forall n = 1\ldots N_t N_r$, where we can write $\pmb{\bar{{h}}}_n = [\pmb{\bar{{h}}}_{n{[1]}}  \cdots \pmb{\bar{{h}}}_{n{[N_P]}}]\in \mathbb{C}^{N_P}$. The task is basically to find the indexes $k$ corresponding to the $q$ most likely peaks $|\pmb{\bar{{h}}}_{n[k]}|$ over all antenna pairs, which are defined as 
%
\begin{equation}
\label{eq:ik}
    k \in \left\{ \widehat{i}_1, \cdots, \widehat{i}_q \right\},
\end{equation}
and are found by analyzing the peaks of all $N_t N_r$ antenna pairs.
%whose details are omitted due to the lack of space.
  Note that a \textcolor{black}{coarse estimation} of the \ac{ToAs} is obtained by scaling the indices as 
    %\begin{equation}
     $ \widehat{\tau}_k =  \frac{1}{N_P \Delta_f} \widehat{i}_k,$
    %\end{equation}
$\forall k = 1 \ldots q$. 
Two extreme cases are discussed:
\textit{(i)} In the well-separated case, i.e. when all $i_k$s are distinct ($i_1 \neq i_2 \neq \ldots \neq i_q$), it is easily verified that the $i_k$-th row of $\pmb{F}^H \pmb{\bar{{H}}}^T$ is the only row that contains information about $\theta_k$ and $\phi_k$. 
\textit{(ii)} In the other extreme case where all $\tau_k$'s arrive within the same time index, which leads to $i \triangleq i_1 =  i_2 = \ldots = i_q$, then the $i^{th}$ row $\pmb{F}^H \pmb{\bar{{H}}}^T$ is the only row that contains all \ac{AoA} and \ac{AoD} information. 
To accommodate for the worst-case, we describe an algorithm able to discriminate between all the \ac{AoAs} and \ac{AoDs}. In that case, the $i^{th}$ row of $\pmb{F}^H \pmb{\bar{{H}}}^T$ can be reshaped as 
\begin{equation}
	\label{eq:system-model}
\textcolor{black}{	\widehat{\pmb{H}} 
	=
	\res_{N_r,N_t} [ (\pmb{F}^H)_{[i,:]} \pmb{\bar{{H}}}^T ]
	=
	\pmb{A}_r(\pmb{\Theta})
	\pmb{G}
	\pmb{A}_t^T(\pmb{\Phi}) + \widehat{\pmb{W}}.}
\end{equation}
\textcolor{black}{Note that $\widehat{\pmb{W}} = \res_{N_r,N_t} [ (\pmb{F}^H)_{[i,:]} \pmb{\bar{{W}}}^T ]$ and $ \pmb{\bar{{W}}}$ is defined in \eqref{eq:Wbar-definition}.}Note that for the well-separated case, the $i_k$-th row of $\pmb{F}^H \pmb{\bar{{H}}}^T$ would be \textcolor{black}{$\alpha_k g_t(\phi_k)g_r(\theta_k) \pmb{a}_r(\theta_k)\pmb{a}^T_t(\phi_k) +  \widehat{\pmb{W}}_k$}, i.e. a special case of \eqref{eq:system-model}. An intermediate case can also be discussed where some paths, say $p < q$, arrive within the same bin. In that case, the corresponding row is also cast as \eqref{eq:system-model}, however, the dimension $q$ would be replaced with $p$. 
\textcolor{black}{Note that both, the channel estimation and the coarse timing estimation, are indeed classical approaches in the context of communications. 
For instance, the channel estimation procedure corresponds to the well-established least-squares.
In addition, the coarse timing estimation process can be used for synchronizing communication frames.
Based on this, the primary objective is to demonstrate that sensing can be effectively achieved through the application of familiar preprocessing techniques. 
These techniques are widely recognized within the research community and are characterized by their ease of implementation.}
\section{The Maximum Likelihood Estimator}
\label{sec:MLE}
In this section, we describe the deterministic \ac{MLE} criterion for the observed data in \eqref{eq:system-model-obs}. 
The deterministic \ac{MLE} regards the sample functions as unknown deterministic sequences, rather than random processes. To this end, we can express the joint density function of the data as follows
\begin{equation*}
\label{eq:likelihood}
\begin{split}
	f(\pmb{\mathcal{Y}})
	=&
	\prod_{n=1}^{N_P}
	\prod_{k=1}^{K_P}
	\frac{1}{\pi \det(\sigma^2 \pmb{I})}  \exp 
	\Big(
	-\frac{1}{\sigma^2}
	\big\Vert 
	\pmb{y}_{n,k}
	-
	\pmb{H}_n
	\pmb{s}_{n,k}
	\big\Vert^2
	\Big),
\end{split}
\end{equation*}
where the $k^{th}$ column of $\pmb{\mathcal{Y}}$ is $ \pmb{\mathcal{Y}}_{[:,k]} 
    =
    \begin{bmatrix}
        \pmb{y}_{1,k} &
        \hdots
        &
        \pmb{y}_{N_P,k}
    \end{bmatrix}^T$
Note that $f(\pmb{\mathcal{Y}})$ is conditioned over the pilots $\pmb{s}_{n,k}$, $\sigma^2$, \textcolor{black}{$\breve{\pmb{\alpha}}=[\breve{\alpha}_0 \ldots \breve{\alpha}_q]$}, $\pmb{\Theta}$, $\pmb{\Phi}$ and this dependency has been omitted for sake of compact notation. We express the log-likelihood as
\begin{equation*}
\label{eq:log-likelihood}
\begin{split}
	\mathcal{L} &\triangleq \log f(\pmb{\mathcal{Y}})
        =
        g(\sigma^2)
        -
        \frac{1}{\sigma^2}
	\sum\limits_{n=1}^{N_P}
	\sum\limits_{k=1}^{K_P}
	\big\Vert 
	\pmb{y}_{n,k}
	-
	\pmb{H}_n
	\pmb{s}_{n,k}
	\big\Vert^2,
\end{split}
\end{equation*}
where $g$ is only a function of $\sigma^2$. As we are interested in the sensing parameters, we can re-write the \ac{MLE} as%in a very compact way as
\begin{equation}
\label{eq:MLE-criterion}
    \argmin_{\breve{\pmb{\alpha}},\pmb{\Theta},\pmb{\Phi},\pmb{\tau}}
    \Big\Vert 
    \pmb{\mathcal{Y}}
    -
    [ \pmb{I}_{N_P} \otimes \pmb{A}_r(\pmb{\Theta})\pmb{G} ]
    \pmb{D}(\pmb{\tau})
    [ \pmb{I}_{N_P} \otimes \pmb{A}_t^T(\pmb{\Phi}) ]
    \pmb{\mathcal{S}}
    \Big\Vert^2
\end{equation}
where the $k^{th}$ column of $\pmb{\mathcal{S}}$ is 
$ \pmb{\mathcal{S}}_{[:,k]} 
    =
    \begin{bmatrix}
        \pmb{s}_{1,k} &
        \hdots
        &
        \pmb{s}_{N_P,k}
    \end{bmatrix}^T$. Also, $\pmb{D}(\pmb{\tau}) = \diag\big(\pmb{D}_1(\pmb{\tau}), \ldots \pmb{D}_{N_P}(\pmb{\tau}) \big)$.
    Note that to obtain the \ac{AoA} and \ac{AoD} information, one has to exhaustively solve the multi-dimensional optimization problem in \eqref{eq:MLE-criterion}, which involves a $5q-$dimensional search. Therefore, we resort to machine learning and parameterized methods to estimate the sensing parameters. 
    More specifically, a grid search on the \ac{MLE} criterion in \eqref{eq:MLE-criterion} would cost $\mathcal{O}(G_\tau^q G_\theta^q G_\phi^q G_\alpha^{2q}.(N_rN_P q^2 N_P^3N_t + N_rN_P^2N_PK_P))$, where  $G_\tau,G_\theta,G_\phi,G_\alpha$ are the grid sizes of the \ac{ToA}, \ac{AoA}, \ac{AoD} and path gains, respectively. 

\section{Machine Learning-based AoA and AoD Estimation}\label{sec:ML_method}
While most of the reviewed literature \cite{surveySa,survey17,survey20} uses the receive matrix\footnote{In our case it corresponds to $\pmb{Y}_n$ containing all \ac{OFDM} symbols.}, the known pilot information $\pmb{S}_{P,n}$ or their covariance matrix as input, we exploit here the estimated channel matrix to extract from the IFFT over its rows a matrix $\widehat{\pmb{H}_{k}}$ of reduced size $N_t \times N_r$ containing all the information about $\theta_k$ and $\phi_k$,  as described in Section~\ref{subsec:coarse-timing-estimation}. As for the deep network design, a straightforward approach is to use an \ac{MLP} and preprocess the complex-valued input to an adequate input tensor, e.g, by concatenating the real, imaginary and complex argument parts of the input, as done in \cite{survey33}, i.e.
%
% \hl{I see the input data as redundant, i.e. the input is using real and imaginary, which are enough to conclude the argument (i.e. $\arg \left\{\widehat{\pmb{H}_{k}}\right\} $). Why do we do this?, They don't explicit say why in the paper. but there is no harm of adding additional features if the model can exploit them directly}
%
\begin{equation}
    \mathfrak{\widehat{\pmb{I}}}_{k}  = \left[\Im\left\{\widehat{\pmb{H}_{k}}\right\} ; \arg \left\{\widehat{\pmb{H}_{k}}\right\} ; \mathfrak{R}\left\{\widehat{\pmb{H}_{k}}\right\}\right] \in \mathbb{R}^{N_t \times N_r \times 3} .
\end{equation}
However, in our work, we leverage complex-valued \acp{NN} \cite{trabelsi2018deep}, inspired by recent findings indicating that complex numbers possess a richer representational capacity and attractive properties. This choice is particularly relevant due to the inherent complex-value operations involved in modeling communication systems. Hence, we use complex convolutional and linear layers as fundamental building blocks for our networks. Indeed, the weights of the linear layers and convolution filters are specifically represented as complex matrices $\pmb{W} = \pmb{W}^r + j \pmb{W}^i$. This approach allows us to effectively leverage the inherent complex-valued operations within these layers, enabling more expressive and accurate modeling capabilities. For example, when performing the equivalent conventional real-valued convolution in the complex domain with a complex vector $\pmb{h}=\pmb{x}+j \pmb{y}$, we have
\begin{equation}
    \pmb{W} * \pmb{h}=(\pmb{W}^r * \pmb{x}-\pmb{W}^i * \pmb{y})+j(\pmb{W}^i * \pmb{x}+\pmb{W}^r * \mathbf{y}) .
\end{equation}
This is due to the distributive property of the convolution operator. Using matrix notation to depict the real and imaginary components of the convolution operation, we have
\begin{equation}
\left[\begin{array}{l}
\Re(\pmb{W} * \pmb{h}) \\
\Im(\pmb{W} * \pmb{h})
\end{array}\right]=\left[\begin{array}{rr}
\pmb{W}^r & -\pmb{W}^i \\
\pmb{W}^i & \pmb{W}^r
\end{array}\right] *\left[\begin{array}{l}
\pmb{x} \\
\pmb{y}
\end{array}\right].
\end{equation}
Similarly, complex linear layers can be constructed using two real-valued linear ones due to the distributive property of the multiplication operator. We illustrate in Fig.~\ref{fig:NN} an example of a complex-valued \ac{NN} where the weights and bias parameters of the $k$-th layer are $\pmb{W}_k =\pmb{W}_k^r+j \pmb{W}_k^i$ and $\pmb{b}_k=$ $\pmb{b}_k^r+j \pmb{b}_k^i$, respectively. The response of this layer to an input formed from the previous layer as $\pmb{z}^{k-1}=\pmb{x}^{k-1}+j \pmb{y}^{k-1}$ is given by:
\begin{equation}
\small
\pmb{z}^k= \left(\pmb{W}_k^r \pmb{x}^{k-1}-\pmb{W}_k^i\pmb{y}^{k-1}+\pmb{b}_k^r \right)+j\left(\pmb{W}_k^r \pmb{y}^{k-1}+\pmb{W}_k^i \pmb{x}^{k-1}+\pmb{b}_k^i\right).
\end{equation}
Moreover, numerous activation functions have been proposed in the existing literature to handle complex-valued representations. However, in our study, we specifically employ the complex rectified linear unit (or $\mathbb{C R} e \mathbb{L U}$). This activation function operates independently on the real and imaginary components of each complex neuron, i.e
\begin{equation}   
\mathbb{C R} e \mathbb{L U}(\pmb{z}^{k})=\operatorname{ReLU}(\Re(\pmb{z}^{k}))+j \operatorname{ReLU}(\Im(\pmb{z}^{k})) .
\end{equation}
It is important to note that there is no need to constrain the network to holomorphic functions, as it was demonstrated that ensuring differentiability of the objective function and activation functions with respect to both the real and imaginary components is a sufficient condition \cite{yoshida}.
%
%In terms of generating training and evaluation data, we use the radar signal model to generate simulated data that encompasses diverse scenarios with randomly positioned targets, resulting in variations in AoA and AoD. 
%Subsequently, the received vector and known pilot information are exploited to compute the channel estimate. 
%Furthermore, the coarse Timing estimation is performed to compute the reduced input matrices corresponding to each of the individual targets. It is worth noting that these procedures are specifically applicable to scenarios with well-separated targets. 
The training data is generated according to the model of Section~\ref{sec:system_model}.
In the case of well-separated targets, the architectures used consist of three hidden layers and a final output layer comprising of $N_{out}=2$ neurons corresponding to the predictions for both \ac{AoA} and \ac{AoD}. The input is a flattened vector of the input  matrix of size $S_{inp} =  N_t N_r$. The three hidden layers are mapping the input to the following latent dimensions $ \left[ \lfloor \frac{S_{inp}}{2} \rfloor, \lfloor \frac{S_{inp}}{4} \rfloor, \lfloor \frac{S_{inp}}{8} \rfloor \right] $. Indeed, a complex linear layer mapping $\mathcal{S}_{i}$ to $\mathcal{S}_{o}$ with a $\mathbb{C R} e \mathbb{L U}$ activation function requires at total $4(\mathcal{S}_{i} \cdot \mathcal{S}_{o}) + 2 \mathcal{S}_{i }$ multiplications and  $4\mathcal{S}_{o} + 3\mathcal{S}_{i}$ additions. Consequently, we have the following total number of operations by summing the contributions of all the different layers for multiplications and additions
\begin{equation}
  \small
  \label{equ:MLPComplexitymul}
  \begin{split}
  T_{\tt{mul}}
     &= 4 \left(  S_{inp} \lfloor \frac{S_{inp}}{2} \rfloor + \lfloor \frac{S_{inp}}{2} \rfloor \lfloor \frac{S_{inp}}{4} \rfloor + \lfloor \frac{S_{inp}}{4} \rfloor \lfloor \frac{S_{inp}}{8} \rfloor \right.\\
     &+ \left. 2 \lfloor \frac{S_{inp}}{8} \rfloor  \right) + 2 \left(S_{inp} + \lfloor \frac{S_{inp}}{2} \rfloor + \lfloor \frac{S_{inp}}{4} \rfloor + 2 N_{out} \right),  
   \end{split}
\end{equation}
\begin{equation}
  \small
  \label{equ:MLPComplexityadd}
  \begin{split}
  T_{\tt{add}}
     &= 3  S_{inp} + 7 \left( \lfloor \frac{S_{inp}}{2} \rfloor + \lfloor \frac{S_{inp}}{4} \rfloor \right) + 4 \lfloor \frac{S_{inp}}{8} \rfloor + 8 N_{out} .
    \end{split}
\end{equation}
For the objective function, the \ac{MSE} for the AoA is given as
\begin{equation}
\mathrm{MSE}_{AoA}=\frac{1}{q E} \sum_{e=1}^E \sum_{k=1}^q\left(\widehat{\theta}_k^{\operatorname{tar}}(e)-\theta_{k}(e)\right)^2,
\end{equation}
where the same formula is used to compute the \ac{MSE} for \ac{AoD} using $\{ \widehat{\phi}_k^{\operatorname{tar}}(e) \}_{k,e}$.
During the training process, the \ac{MSE} between the estimates and true targets is minimized. This minimization allows tuning the network weights based on the gradient of the objective function. To ensure the convergence of the model, the angles ${ \theta_i, \phi_i } , \forall i $ are sorted before computing the error, as it is necessary for the target angles to follow a deterministic order.

% scriptsize
\section{Parameterized $2$D Algorithm}\label{sec:benchmarking_algorithm}
\subsection{Algorithmic Description}
\label{sec:algorithmic-description-2D}
Since implementing the maximum likelihood estimation is not feasible according to Section \ref{sec:MLE}, in this section we present a parameterized method that has knowledge of the system model of Section~\ref{sec:system_model}, which will be used as a benchmark for the ML-based approach presented in Section~\ref{sec:ML_method}.

Let $M_t \leq N_t$ and $M_r \leq N_r$ be the sub-array sizes. We perform  a data transformation by exploiting the structure of \ac{ULA} array configuration. To this end, we form $\widehat{\pmb{H}} $ as such
\begin{equation}
\label{step1-1}
	\widehat{\pmb{\mathcal{H}}}
	=
	\begin{bmatrix}
		\widehat{\pmb{\mathcal{H}}}_1 
		& 
		\widehat{\pmb{\mathcal{H}}}_2
		&
		\hdots
		&
		\widehat{\pmb{\mathcal{H}}}_{K_t	} \\
		\widehat{\pmb{\mathcal{H}}}_2 
		& 
		\widehat{\pmb{\mathcal{H}}}_3
		&
		\hdots
		&
		\widehat{\pmb{\mathcal{H}}}_{K_t + 1} \\
		\vdots
		& 
		\vdots
		&
		\ddots
		&
		\vdots \\
		\widehat{\pmb{\mathcal{H}}}_{M_t} 
		& 
		\widehat{\pmb{\mathcal{H}}}_{M_t + 1}
		&
		\hdots
		&
		\widehat{\pmb{\mathcal{H}}}_{N_t} 
	\end{bmatrix},
\end{equation}
where $K_t \triangleq N_t - M_t + 1$ is the number of sub-arrays formed by Tx array. Each Hankel matrix $\widehat{\pmb{\mathcal{H}}}_{i} \in \mathbb{C}^{M_r \times K_r}$ is formed as 
% \begin{equation}
% \label{step1-2}
% 	\widehat{\pmb{\mathcal{H}}}_{i}
% 	=
% 	\begin{bmatrix}
% 		\widehat{{H}}_{i,1}
% 		& 
% 		\widehat{{H}}_{i,2}
% 		&
% 		\hdots
% 		&
% 		\widehat{{H}}_{i,K_r} \\
% 		\widehat{{H}}_{i,2}
% 		& 
% 		\widehat{{H}}_{i,3}
% 		&
% 		\hdots
% 		&
% 		\widehat{{H}}_{i,K_r + 1} \\
% 		\vdots
% 		& 
% 		\vdots
% 		&
% 		\ddots
% 		&
% 		\vdots \\
% 		\widehat{{H}}_{i,M_r}
% 		& 
% 		\widehat{{H}}_{i,M_r+1}
% 		&
% 		\hdots
% 		&
% 		\widehat{{H}}_{i,N_r} \\
% 	\end{bmatrix},
% \end{equation}
\begin{equation}
\label{step1-2}
	[\widehat{\pmb{\mathcal{H}}}_{i}]_{m,n}
	=
        \widehat{{H}}_{i,m+n-1},
\end{equation}
where $i = 1 \ldots N_t$ and $K_r = N_r - M_r + 1$ represents the number of sub-arrays formed by the receive array. Thanks to this manipulation, we can re-write equation \eqref{eq:system-model} as follows
\begin{equation}
	\label{eq:system-model-2}
	\widehat{\pmb{\mathcal{H}}}
	=
	\pmb{\mathcal{A}}_{M_r,M_t}(\pmb{\Theta},\pmb{\Phi})
	\pmb{G}
	\pmb{\mathcal{A}}_{K_r,K_t}^T(\pmb{\Theta},\pmb{\Phi}) + \widetilde{\pmb{\mathcal{W}}},
\end{equation}
\begin{equation}
\text{where} \quad	\pmb{\mathcal{A}}_{n,m}(\pmb{\Theta},\pmb{\Phi})
	=
	\begin{bmatrix}
		\pmb{A}_r(\pmb{\Theta})_{[1:n,:]}  \\
		\pmb{A}_r(\pmb{\Theta})_{[1:n,:]} \pmb{D}_\phi(\pmb{\Phi}) \\
		\vdots \\
		\pmb{A}_r(\pmb{\Theta})_{[1:n,:]} \pmb{D}^{m-1}(\pmb{\Phi})
	\end{bmatrix},
\end{equation}
with $\pmb{D}_\phi(\pmb{\Phi}) = \diag [ \pmb{a}_1(\phi_1) \ldots\pmb{a}_1(\phi_q) )]$ and $\widetilde{\pmb{\mathcal{W}}}$ contains entries of $\widetilde{\pmb{W}}$.
In contrast to the model in equation \eqref{eq:system-model}, \eqref{eq:system-model-2} includes interaction between both \ac{AoA} and \ac{AoD} within the left and right sub-spaces, due to the inflated dimensions introduced by the sub-arrays at both transmit and receive ends. Now, given \eqref{eq:system-model-2}, two overlapping matrices can be extracted from $\widehat{\pmb{\mathcal{H}}}$, 
% \begin{equation}
% 	\label{eq:step2_1}
% \begin{split}
% 	\widehat{\pmb{\mathcal{H}}}^{(1)} 
% 	\triangleq
% 	\widehat{\pmb{\mathcal{H}}}_{[:,1:K_r(K_t-1)]} = 
% 	\pmb{\mathcal{A}}_{M_r,M_t}(\pmb{\Theta},\pmb{\Phi})
% 	\pmb{G}
% 	\pmb{\Pi}^T
% 	+
% 	\widehat{\pmb{\mathcal{W}}}^{(1)},
% \end{split}
% \end{equation}
% and
% \begin{equation}
% 	\label{eq:step2_2}
% \begin{split}
% 	\widehat{\pmb{\mathcal{H}}}^{(2)} 
% 	\triangleq
% 	\widehat{\pmb{\mathcal{H}}}_{[:,(K_r+1):K_rK_t]} = 
% 	\pmb{\mathcal{A}}_{M_r,M_t}(\pmb{\Theta},\pmb{\Phi})
% 	\pmb{G}
% 	\pmb{D}(\pmb{\Phi})
% 	\pmb{\Pi}^T
% 	+
% 	\widehat{\pmb{\mathcal{W}}}^{(2)},
% \end{split}
% \end{equation}
\begin{align}
\label{eq:step2_1}
	\widehat{\pmb{\mathcal{H}}}^{(1)} 
	&\triangleq
	\widehat{\pmb{\mathcal{H}}}_{[:,1:K_r(K_t-1)]} = 
	\pmb{\mathcal{A}}_{M_r,M_t}(\pmb{\Theta},\pmb{\Phi})
	\pmb{G}
	\pmb{\Pi}^T
	+
	\widehat{\pmb{\mathcal{W}}}^{(1)}, \\
 \label{eq:step2_2}
	\widehat{\pmb{\mathcal{H}}}^{(2)} 
	&\triangleq
	\widehat{\pmb{\mathcal{H}}}_{[:,K_r+1:K_rK_t]} = 
	\pmb{\mathcal{A}}_{M_r,M_t}(\pmb{\Theta},\pmb{\Phi})
	\pmb{G}
	\pmb{D}_\phi(\pmb{\Phi})
	\pmb{\Pi}^T
	+
	\widehat{\pmb{\mathcal{W}}}^{(2)}, 
\end{align}
Interestingly, both matrices can be exploited to compute $\phi_1 \ldots \phi_q$. Note that for the specific matrix $\widehat{\pmb{\mathcal{H}}}_\gamma \triangleq \widehat{\pmb{\mathcal{H}}}^{(2)}  - \gamma \widehat{\pmb{\mathcal{H}}}^{(1)} $, we have that
\begin{equation}
\begin{split}
	 \widehat{\pmb{\mathcal{H}}}_\gamma
	=
	\pmb{\mathcal{A}}_{M_r,M_t}(\pmb{\Theta},\pmb{\Phi})
	\pmb{G}
	\Big( 
	\pmb{D}_\phi(\pmb{\Phi})
	-
	\gamma
	\pmb{I}
	\Big)
	\pmb{\Pi}^T
	+
	\widehat{\pmb{\mathcal{W}}}_\gamma ,
 \end{split}
\end{equation}
where $\widehat{\pmb{\mathcal{W}}}_\gamma = \widehat{\pmb{\mathcal{W}}}^{(2)}-\gamma \widehat{\pmb{\mathcal{W}}}^{(1)}$. In the absence of noise, and given that $\pmb{\Pi}$ and $\pmb{\mathcal{A}}_{M_r,M_t}(\pmb{\Theta},\pmb{\Phi})$ are full-column rank, the rank of $\widehat{\pmb{\mathcal{H}}}_\gamma$ drops from $q$ to $q-1$ at $\gamma = \pmb{a}_1(\phi_i)$, $\forall i = 1 \ldots q$. Based on this, one approach can be to perform an exhaustive search over $\phi$ as $\gamma = \pmb{a}_1(\phi)$ and evaluate the $q^{th}$ largest singular value of $\widehat{\pmb{\mathcal{H}}}_\gamma$, subsequently. Then the $q$ minima of the resulting spectrum provide the \ac{AoD} estimates. However, such an approach is computationally exhaustive as a \ac{SVD} is required per point $\phi$. Instead, we perform a single \ac{SVD} by first obtaining $\widehat{\pmb{\mathcal{H}}}^{(1)}  = \pmb{U}\pmb{\Sigma}\pmb{V}^H,$ where $\pmb{U},\pmb{V}$ are the left/right singular vectors of $\widehat{\pmb{\mathcal{H}}}^{(1)}$ and $\pmb{\Sigma}$ contains the singular values of $\widehat{\pmb{\mathcal{H}}}^{(1)}$ in decreasing order. Then, we truncate the \ac{SVD} by first truncating $\pmb{\bar{\Sigma}} \in \mathbb{C}^{q \times q}$ by picking the upper-left $q \times q$ sub-matrix of $\pmb{\Sigma}$. Similarly, $\pmb{\bar{U}},\pmb{\bar{V}}$ are the associated singular vectors of $\pmb{\bar{\Sigma}} \in \mathbb{C}^{q \times q}$ corresponding to the $q$ strongest singular values. Next, we compute the eigenvalues of the matrix 
\begin{equation}
\label{eq:step4}
	\pmb{T} = 
	\pmb{\bar{\Sigma}}^{-1}
	\pmb{\bar{U}}^H
	\widehat{\pmb{\mathcal{H}}}^{(2)}
	\pmb{\bar{V}},
\end{equation}
which are denoted as $\gamma_1 \ldots \gamma_q$. These eigenvalues are estimates of $\pmb{a}_1(\hat{\phi}_i)$, $\forall i $. Therefore, we can extract $\hat{\phi}_i$, $\forall i$, as follows
\begin{equation}
\label{eq:step6}
	\hat{\phi}_i = -\sin ^{-1} \Big( \frac{\lambda \arg(  \gamma_i ) }{2 \pi d_t} \Big) , \forall i = 1 \ldots q .
\end{equation}  
Following the \ac{AoD} estimates, we turn our attention to \ac{AoA} estimation. A two-staged \ac{LS} fit is proposed. The first stage entails obtaining a non-parametrized estimate of the \ac{AoA} manifold via the following \ac{LS} criterion. 
\begin{equation}
\begin{split}
	\label{eq:step7}
	\widehat{\pmb{X}}
	& =
	\argmin_{\pmb{X}}
	\big
	\Vert
	\widehat{\pmb{H}} 
	-
	\pmb{X}	
	\pmb{A}_t^T(\widehat{\pmb{\Phi}}) \big\Vert^2
	 =
	\widehat{\pmb{H}}
	\pmb{A}_t^*(\widehat{\pmb{\Phi}})
	\big(
	\pmb{A}_t^T(\widehat{\pmb{\Phi}})
	\pmb{A}_t^*(\widehat{\pmb{\Phi}})
	\big)^{-1}.
	\end{split}
\end{equation}
The second stage exploits $\widehat{\pmb{X}}$ to obtain a non-parametrized estimate of an un-parametrized version of $\widehat{\pmb{A}_r}$ under a per-column norm constraint on $\pmb{A}_r$. Based on this, we can write
\textcolor{black}{
\begin{equation}
\label{eq:estimate-Ar-alpha-breve}
(\widehat{\pmb{A}_r},\widehat{\breve{\pmb{\alpha}}})
=
\begin{cases}
\argmin_{\pmb{A}_r,\breve{\pmb{\alpha}}}
 &
\big
\Vert
	\widehat{\pmb{X}}
	-
	\pmb{A}_r
	\pmb{G}	
\big\Vert^2, \\
\subjectto &\Vert {\pmb{A}_r}_{[:,i]}\Vert = 1,  \pmb{G} = \diag(\breve{\pmb{\alpha}}).
\end{cases}
\end{equation}
}
But since $\pmb{G}$ is diagonal, the computation is decoupled and hence $\widehat{\pmb{A}_r}$ can be computed on a column-by-column basis as
\textcolor{black}{
\begin{equation}
\label{eq:estimate-Ar-i}
\widehat{\pmb{A}_r}_{[:,i]}
=
\begin{cases}
\argmin_{\pmb{a}_i} &
\big
	\Vert
	\widehat{\pmb{X}}_{[:,i]}
	-
	\breve{\alpha}_i
	\pmb{a}_i	
\big\Vert^2, \\
\subjectto &\Vert \pmb{a}_i \Vert = 1,
\end{cases}
\end{equation}
}
where the solution can be shown to be $	\widehat{\pmb{a}}_{i} = \frac{\widehat{\pmb{X}}_{[:,i]}}{\Vert \widehat{\pmb{X}}_{[:,i]} \Vert}$
% \begin{equation}
% \label{eq:step8}
% 	\widehat{\pmb{a}}_{i} = \frac{\widehat{\pmb{X}}_{[:,i]}}{\Vert \widehat{\pmb{X}}_{[:,i]} \Vert}, \qquad  \forall i = 1 \ldots q
% \end{equation}
and \textcolor{black}{$\vert \widehat{\breve{\alpha}}_i \vert = \frac{\Vert \widehat{\pmb{X}}_{[:,i]} \Vert}{\sqrt{N_r}}$}.
It is worth noting that a pairing/matching  method is not needed because $\widehat{\pmb{a}}_{i}$ is associated with the $i^{th}$ column of $\pmb{A}_t(\widehat{\pmb{\Phi}})$, i.e. $\pmb{a}_{N_t}(\phi_i)$. Finally, given un-parameterized steering vectors, we perform a simple linear regression on the phases of $\pmb{a}_i$ to obtain the \ac{AoA}s. To this end, we have the following
\begin{equation}
\label{eq:step09}
\left(\hat{\theta}_i,\hat{\delta}_i\right)=
\argmin_{\theta_i,\delta_i}
\Big
	\Vert
	\arg( \widehat{\pmb{a}}_{i} )
	-
	\pmb{\Xi}
	\begin{bmatrix}
		\theta_i \\
		\delta_i
	\end{bmatrix}
\Big
	\Vert^2 = \pmb{\Xi}^\dagger \arg(  \widehat{\pmb{a}}_{i}),
\end{equation}
$\forall i = 1\ldots q$, where $\pmb{\Xi} \in \mathbb{R}^{N_r \times 2}$ where the first column contains all integers counting from $1$ to $N_r$ and the second column is all-ones. 
\textcolor{black}{It is worth highlighting that the phase offset $\delta_i$ corresponding to the $i^{th}$ target not only contains the unknown phase of the channel coefficient (which includes the pathloss and the radar cross-section coefficient) $\alpha_i$, but also the unknown phases of the transmit antenna radiation pattern due to the \ac{BS} and the receive antenna radiation pattern at the radar. Therefore, its estimate, hereby denoted as $\hat{\delta}_i$ can be expressed as
\begin{equation}
	\hat{\delta}_i
	=
	\arg
	\big(
	\alpha_i
	\big)
	+
	\arg
	\big(
	g_t(\phi_i)
	\big)
	+
	\arg
	\big(
	g_r(\theta_i)
	\big).
\end{equation}
}
Note that the phases of $\arg(  \pmb{a}_i )$ should be unwrapped to provide a smooth phase linear estimation. We note here that in \eqref{eq:step09}, one can consider replacing $\pmb{a}_i$ with ${\widehat{\pmb{X}}_{[:,i]}}$ since we are using the phases of $\pmb{a}_i$.
A summary of the algorithm is given in \textbf{Algorithm \ref{alg:alg1}}.

\begin{algorithm}[H]
\caption{Sensing \ac{AoA}/\ac{AoD} via Coarse \ac{ToA} Estimates}\label{alg:alg1}
\begin{algorithmic}
\STATE 
\STATE {\textsc{input}: $\pmb{Y}_P,\pmb{S}_P$}
\STATE {\textsc{Channel Estimation}:} 
\STATE \hspace{0.5cm} Obtain $\lbrace \widehat{\pmb{H}}_n \rbrace_{n=1}^{N_P}$ according to equation \eqref{eq:method2_step0}.\\

\STATE {\textsc{Coarse Timing Estimation}:} 
\STATE\hspace{0.5cm}{\tt{1}}) For each $n = 1 \ldots N_tN_r$, get $\widehat{\pmb{h}}_n$ by \ac{IFFT} as in \eqref{eq:IFFT-step}.

\STATE\hspace{0.5cm}{\tt{2}}) Get the $q$ mostly occurred values in as given by \eqref{eq:ik}.

\STATE {\textsc{Sensing Estimation}:} 
\STATE\hspace{0.5cm}{\tt{0}}) Using the $ \widehat{i}_k$-th row of $\pmb{F}^H \pmb{\bar{{H}}}^T$, form $\widehat{\pmb{H}} $ using \eqref{eq:system-model}.

\STATE\hspace{0.5cm}{\tt{1}}) Form $\widehat{\pmb{\mathcal{H}}}$ as described in equations \eqref{step1-1} and \eqref{step1-2}. 
\STATE\hspace{0.5cm}{\tt{2}}) Extract $\widehat{\pmb{\mathcal{H}}}^{(1)}$ and $\widehat{\pmb{\mathcal{H}}}^{(2)}$ as \eqref{eq:step2_1} and \eqref{eq:step2_2}, respectively.
\STATE\hspace{0.5cm}{\tt{3}}) Compute a truncated \ac{SVD} of $\widehat{\pmb{\mathcal{H}}}^{(1)}$ as 
	\begin{equation*}
		\big[ \pmb{\bar{U}}, \pmb{\bar{\Sigma}}, \pmb{\bar{V}}\big] \gets \TSVD_q(\widehat{\pmb{\mathcal{H}}}^{(1)}).
	\end{equation*}
\STATE\hspace{0.5cm}{\tt{4}}) Given $\pmb{\bar{U}}, \pmb{\bar{\Sigma}}, \pmb{\bar{V}},\widehat{\pmb{\mathcal{H}}}^{(1)}$, compute $\pmb{T}$ using \eqref{eq:step4}. 
\STATE\hspace{0.5cm}{\tt{5}}) Get the eigenvalues of $\pmb{T}$, i.e. $\lbrace {\gamma}_i \rbrace_{i=1}^q$.
\STATE\hspace{0.5cm}{\tt{6}}) For the $i^{th}$ eigenvalue, estimate the $i^{th}$ \ac{AoD} via \eqref{eq:step6}.
\STATE\hspace{0.5cm}{\tt{7}}) Perform an \ac{LS}-fit following \eqref{eq:step7} to obtain $\widehat{\pmb{X}}$.
\STATE\hspace{0.5cm}{\tt{8}}) Given $\widehat{\pmb{X}}$, obtain $\lbrace \widehat{\pmb{a}}_i \rbrace_{i=1}^q$ through $	\widehat{\pmb{a}}_{i} = \frac{\widehat{\pmb{X}}_{[:,i]}}{\Vert \widehat{\pmb{X}}_{[:,i]} \Vert}$.
\STATE\hspace{0.5cm}{\tt{9}}) For each $i$, obtain $\hat{\theta}_i$ as advised in \eqref{eq:step09}.
\STATE \textbf{return}  $(\hat{\theta}_1,\hat{\phi}_1) \ldots (\hat{\theta}_q,\hat{\phi}_q)$.
\end{algorithmic}
\end{algorithm}

	\label{eq:step2-1}
%	\widehat{\pmb{\mathcal{H}}}^{(1)} 
\subsection{\textcolor{black}{Required Estimation Overhead \&  Complexity}}
\label{sec:required-estimation-overhead-and-complexity}

\textcolor{black}{%Now, we explain the required estimation overheads along with the computational complexity required per block. These blocks are
The required estimation overheads, along with the computational complexity, required per block is explained as follows:
\begin{itemize}
	\item \textbf{Channel Estimation}
	
	The channel estimation step in \eqref{eq:method2_step0} requires a series of matrix multiplications and an inverse to obtain $\widehat{\pmb{H}}_n$.
	A total of $N_t^2N_PK_P + N_t^3 + N_PK_PN_t^2 + N_rN_PK_PN_t$ multiplications and $N_t(N_PK_P-1)N_t + N_t^3 - 2N_t^2 + N_t + N_PK_P(N_t-1)N_t + N_r(N_PK_P-1)N_t$ additions is required to perform the channel estimation block.
	
	\item \textbf{Coarse Timing Estimation}
	
	This block mainly performs $N_tN_r$ \ac{IFFT} operations, where each consists of $N_P^2$ multiplications and $N_P(N_P-1)$ additions, and hence a total of $N_P^2N_tN_r$ multiplications and $N_P(N_P-1)N_tN_r$ additions. 
	
	\item \textbf{Sensing Estimation}
	
	\begin{itemize}
		\item Step {\tt{0}}, Step {\tt{1}} and Step {\tt{2}} within the sensing estimation block of \textbf{Algorithm \ref{alg:alg1}} do not necessitate any floating-point operations, as we are solely arranging  sub-matrices.
	
		\item Step {\tt{3}} within the sensing estimation block of \textbf{Algorithm \ref{alg:alg1}} performs an \ac{SVD} operation, which can be realized using the Golub–Reinsch algorithm. This costs $4 (M_r M_t)^2 K_r(K_t - 1) + 8M_r M_t(K_r(K_t - 1))^2 + 9 (K_r(K_t - 1))^3$ multiplications and $4 (M_r M_t)^2 K_r(K_t - 1) + 8M_r M_t(K_r(K_t - 1))^2 + 9 (K_r(K_t - 1))^3$ additions \cite{1206682,van1996matrix}.
		
	     \item Step {\tt{4}} requires the computation of $\pmb{T}$ following  \eqref{eq:step4}, which involves the matrix multiplication of $\widehat{\pmb{\mathcal{H}}}^{(2)} \in \mathbb{C}^{M_r M_t \times K_r(K_t - 1)}$ with $\pmb{\bar{V}} \in \mathbb{C}^{K_r(K_t - 1) \times q}$, which costs $M_r M_t K_r(K_t - 1) q$ multiplications and $M_r M_t(K_r(K_t - 1) - 1)q$ additions. 
     Then, we multiply $\pmb{\bar{U}}^H  \in \mathbb{C}^{q \times M_r M_t}$ with $\widehat{\pmb{\mathcal{H}}}^{(2)} \pmb{\bar{V}} \in \mathbb{C}^{M_r M_t \times q}$, which costs $M_r M_t q^2$ multiplications and $ ( M_r M_t-1)q^2$ additions.
     Finally, we multiply $\pmb{\bar{\Sigma}}^{-1} \in \mathbb{C}^{q \times q}$ with $	\pmb{\bar{U}}^H
	\widehat{\pmb{\mathcal{H}}}^{(2)}
	\pmb{\bar{V}} \times \mathbb{C}^{q \times q}$. Since $\pmb{\bar{\Sigma}}^{-1}$ is diagonal, then only $q^2$ multiplication are required. 
    Therefore, we conclude that the computation of $\pmb{T} $ requires $M_r M_t K_r(K_t - 1) q + M_r M_t q^2 + q^2 $ multiplication and $M_r M_t(K_r(K_t - 1) - 1)q + ( M_r M_t-1)q^2 $ additions.
    % total: $M_r M_t K_r(K_t - 1) q + M_r M_t q^2 + q^2 $ multiplication and $M_r M_t(K_r(K_t - 1) - 1)q + ( M_r M_t-1)q^2 $ additions

\item In Step {\tt{5}}, we calculate the $q$ complex eigenvalues of $\pmb{T}$. Since $\pmb{T}$ lacks a specific structure, a QZ decomposition (generalized Schur decomposition) is suitable for generating eigenvalues, which costs $6q^2(q-1)$ multiplications, $6q^2(q-1)$ additions, and $2q(q-1)$ square roots \cite{ward1975combination}. If each square root employs \ac{CORDIC} \cite{523811}, a square root operation costs $2N_{\tt{cord}}$ additions and $1$ multiplication, where $N_{\tt{cord}}$ is the number of iterations for the \ac{CORDIC} algorithm, typically dependent on the output's bit size. Ignoring shifting operations, the overall square roots necessitate $4q(q-1)N_{\tt{cord}}$ additions and $2q(q-1)$ multiplications.
\item In Step {\tt{6}}, in order to estimate the \ac{AoD} via equation \eqref{eq:step6}, it is important to note that two \ac{CORDIC} operations are necessary per eigenvalue. One operation is dedicated to the phase retrieval, while the other is for the inverse-sine operation. Assuming both \ac{CORDIC} algorithms employ the same number of iterations as in the square root computation mentioned earlier (i.e., $N_{\tt{cord}}$), the total number of operations needed for \ac{AoD} estimation is $4N_{\tt{cord}}q$ additions and $3q$ multiplications. It's worth noting that there is an additional multiplication per \ac{AoD} arising from the term $\frac{\lambda}{2 \pi d_t}$.

\item In Step {\tt{7}}, the \ac{LS}-fit step specified in \eqref{eq:step7} involves several computational stages. Initially, the computation of $\pmb{A}_t^T(\widehat{\pmb{\Phi}})\pmb{A}_t^*(\widehat{\pmb{\Phi}})$ requires $q^2 N_t$ multiplications and $q^2 (N_t - 1)$ additions. Subsequently, the inversion of this matrix incurs $q^3$ multiplications and $q^3 - 2q^2 + q$ additions. Following this, the computation of $\pmb{A}_t^*(\widehat{\pmb{\Phi}})\big(\pmb{A}_t^T(\widehat{\pmb{\Phi}})\pmb{A}_t^*(\widehat{\pmb{\Phi}})\big)^{-1}$ necessitates $q^2 N_t$ multiplications and $N_t(q-1)q$ additions. The final step involves obtaining $\widehat{\pmb{X}}$ through the multiplication of $\widehat{\pmb{H}} \in \mathbb{C}^{N_r \times N_t}$ with $\pmb{A}_t^*(\widehat{\pmb{\Phi}})\big(\pmb{A}_t^T(\widehat{\pmb{\Phi}})\pmb{A}_t^*(\widehat{\pmb{\Phi}})\big)^{-1} \in \mathbb{C}^{N_t \times q}$, incurring $qN_rN_t$ multiplications and $N_r(N_t-1)q$ additions. Consequently, the overall \ac{LS}-fit operation includes $q^2 N_t + q^3 + q^2 N_t + qN_rN_t$ multiplications and $q^2 (N_t - 1) + q^3 - 2q^2 + q + N_t(q-1)q + N_r(N_t-1)q$ additions.

\item In Step {\tt{9}}, which involves the \ac{AoA} computation, we can make use of the structure of $\pmb{\Xi}$, which would then lead to a total cost of $q(N_r + 2)$ multiplications and $q(2N_r - 1)$ additions.
	\end{itemize}

\end{itemize}

}

\textcolor{black}{%\begin{table*}[!t]
%\caption{Required Estimation Overhead for Sensing AoA/AoD\label{tab:table1}}
%\centering
%{
%\begin{tabular}{|c||c||c|}
%\hline
%\textbf{Block} & \textbf{Multiplications} & \textbf{Additions} \\
%\hline
%Channel Estimation 
%& 
%$ N_t^3 + 2N_PK_PN_t^2 + N_rN_PK_PN_t$ 
%& 
%$N_t^3 - 2N_t^2 + N_t + 
%N_t(N_PK_P-1)N_t + 
%N_PK_P(N_t-1)N_t + 
%N_r(N_PK_P-1)N_t$\\
%\hline
%Coarse Timing Estimation & $N_P^2N_tN_r$  & $N_P(N_P-1)N_tN_r$ 
% \\
%\hline
%Step {\tt{0}}, {\tt{1}}, {\tt{2}} of Sensing Estimation & $\mathcal{O}(1)$ & $\mathcal{O}(1)$   \\
%\hline
%\ac{SVD} computation (Step {\tt{3}}) &  $4 (M_r M_t)^2 K_r(K_t - 1) + 8M_r M_t(K_r(K_t - 1))^2 + 9 (K_r(K_t - 1))^3$  & $4 (M_r M_t)^2 K_r(K_t - 1) + 8M_r M_t(K_r(K_t - 1))^2 + 9 (K_r(K_t - 1))^3$  \\
%\hline
%$\pmb{T}$ computation (Step {\tt{4}}) & $M_r M_t K_r(K_t - 1) q + M_r M_t q^2 + q^2 $ &  $M_r M_t(K_r(K_t - 1) - 1)q + ( M_r M_t-1)q^2 $ additions \\
%\hline
%EVD computation (Step {\tt{5}})  & $2q(q-1)$  & $4q(q-1)N_{\tt{cord}}$   \\
%\hline
%\ac{AoD} Estimation (Step {\tt{6}})  & $3q$ & $4N_{\tt{cord}}q$   \\
%\hline
%LS-fit (Step {\tt{7}}) & $q^2 N_t + q^3 + q^2 N_t + qN_rN_t$ &$q^2 (N_t - 1) + q^3 - 2q^2 + q + N_t(q-1)q + N_r(N_t-1)q$   \\
%\hline
%Step {\tt{8}} & $\mathcal{O}(1)$ & $\mathcal{O}(1)$ \\
%\hline
%\ac{AoA} Estimation (Step {\tt{9}}) & $q(N_r + 2)$ &  $q(2N_r - 1)$   \\
%\hline 
%\end{tabular}
%}
%\end{table*}

\begin{table*}[!t]
\caption{\textcolor{black}{Required Estimation Overhead (Multiplications) for Sensing AoA/AoD\label{tab:table1}}}
\centering
{
\color{black}
\begin{tabular}{|c||c|}
\hline
\textbf{Block} & \textbf{Complex Multiplications}\\
\hline
Channel Estimation 
& 
$ N_t^3 + 2N_PK_PN_t^2 + N_rN_PK_PN_t$ 
\\
\hline
Coarse Timing Estimation & $N_P^2N_tN_r$ 
 \\
\hline
Step {\tt{0}}, {\tt{1}}, {\tt{2}} of Sensing Estimation & $\mathcal{O}(1)$    \\
\hline
\ac{SVD} computation (Step {\tt{3}}) &  $4 (M_r M_t)^2 K_r(K_t - 1) + 8M_r M_t(K_r(K_t - 1))^2 + 9 (K_r(K_t - 1))^3$   \\
\hline
$\pmb{T}$ computation (Step {\tt{4}}) & $M_r M_t K_r(K_t - 1) q + M_r M_t q^2 + q^2 $  \\
\hline
EVD computation (Step {\tt{5}})  & $2q(q-1)$  \\
\hline
\ac{AoD} Estimation (Step {\tt{6}})  & $3q$   \\
\hline
LS-fit (Step {\tt{7}}) & $q^2 N_t + q^3 + q^2 N_t + qN_rN_t$    \\
\hline
Step {\tt{8}} & $\mathcal{O}(1)$  \\
\hline
\ac{AoA} Estimation (Step {\tt{9}}) & $q(N_r + 2)$  \\
\hline 
\end{tabular}
}
\end{table*}

\begin{table*}[!t]
\caption{\textcolor{black}{Required Estimation Overhead (Additions) for Sensing AoA/AoD\label{tab:table2}}}
\centering
{
\color{black}
\begin{tabular}{|c||c|}
\hline
\textbf{Block}  & \textbf{Complex Additions} \\
\hline
Channel Estimation 
& 
%%$
%N_t^3 - 2N_t^2 + N_t + N_t(N_PK_P-1)N_t + N_PK_P(N_t-1)N_t + 
%%N_r(N_PK_P-1)N_t
%%$
$\!\begin{aligned}[t]
    N_t^3 &- 2N_t^2 + N_t  N_PK_P(N_t-1)N_t \\
    &+ N_t(N_PK_P-1)N_t + N_r(N_PK_P-1)N_t
    \end{aligned}$
\\
\hline
Coarse Timing Estimation   & $N_P(N_P-1)N_tN_r$ 
 \\
\hline
Step {\tt{0}}, {\tt{1}}, {\tt{2}} of Sensing Estimation  & $\mathcal{O}(1)$   \\
\hline
\ac{SVD} computation (Step {\tt{3}}) & $4 (M_r M_t)^2 K_r(K_t - 1) + 8M_r M_t(K_r(K_t - 1))^2 + 9 (K_r(K_t - 1))^3$  \\
\hline
$\pmb{T}$ computation (Step {\tt{4}}) &  $M_r M_t(K_r(K_t - 1) - 1)q + ( M_r M_t-1)q^2 $  \\
\hline
EVD computation (Step {\tt{5}})  & $4q(q-1)N_{\tt{cord}}$   \\
\hline
\ac{AoD} Estimation (Step {\tt{6}})  & $4N_{\tt{cord}}q$   \\
\hline
LS-fit (Step {\tt{7}}) &$q^2 (N_t - 1) + q^3 - 2q^2 + q + N_t(q-1)q + N_r(N_t-1)q$   \\
\hline
Step {\tt{8}}  & $\mathcal{O}(1)$ \\
\hline
\ac{AoA} Estimation (Step {\tt{9}}) &  $q(2N_r - 1)$   \\
\hline 
\end{tabular}
}
\end{table*}

}

The computational complexity, excluding channel estimation, of \textbf{Algorithm \ref{alg:alg1}} in terms of the total number of additional and multiplications is calculated by summing up all additions and multiplications of the above sub-blocks. To this end, we have the following total number of operations

\begin{equation*}
   \begin{split}
    T_{\tt{add}}
    &=
    4 (M_r M_t)^2 K_r(K_t - 1) + 8M_r M_t(K_r(K_t - 1))^2  \\
    &+ 9 (K_r(K_t - 1))^3 + M_r M_t(K_r(K_t - 1) - 1)q \\
    &+ ( M_r M_t-1)q^2  + 6q^2(q-1) + 4q(q-1)N_{\tt{cord}} \\
    & + 4N_{\tt{cord}}q + q^2 (N_t - 1)  + q^3 - 2q^2 + q + N_t(q-1)q  \\
    &+ N_r(N_t-1)q + q(2N_r - 1)
   \end{split}
\end{equation*}
%and
\begin{equation*}
   \begin{split}
    T_{\tt{mul}}
   & =
 4 (M_r M_t)^2 K_r(K_t - 1) + 8M_r M_t(K_r(K_t - 1))^2  \\
   &+ 9 (K_r(K_t - 1))^3 + M_r M_t K_r(K_t - 1) q + M_r M_t q^2  \\
   &+ q^2 + q^2(q-1) + 2q(q-1) + 3q + q^2 N_t + q^3 \\
   &+ q^2 N_t + qN_rN_t + q(N_r + 2)
   \end{split}
\end{equation*}
where $ T_{\tt{add}}$ and $ T_{\tt{mul}}$ are the total number of additional and multiplications, respectively.
Comparing the complexity of the $2$D parameterized algorithm with that of \ac{MLE}, we first express the complexity in the order of $\mathcal{O}(N_r^3 N_t^3 q + N_r N_t q^2 + q^3)$, where we have upper-bounded both $M_r,K_r$ by $N_r$ and $M_t,K_t$ by $N_t$. Defining the complexity gain of the parameterized method with respect to \ac{MLE} as $S$, we can say
\begin{equation}
\label{eq:S-factor}
\textcolor{black}{    S
    =
    \frac{\mathcal{O}(G_\tau^q G_\theta^q G_\phi^q G_\alpha^{2q}.(N_rN_P q^2 N_P^3N_t + N_rN_P^2N_PK_P))}{T_{\tt{add}} + T_{\tt{mul}}}.}
\end{equation}

\begin{figure}[!t]
\centering
\includegraphics[width=3.5in]{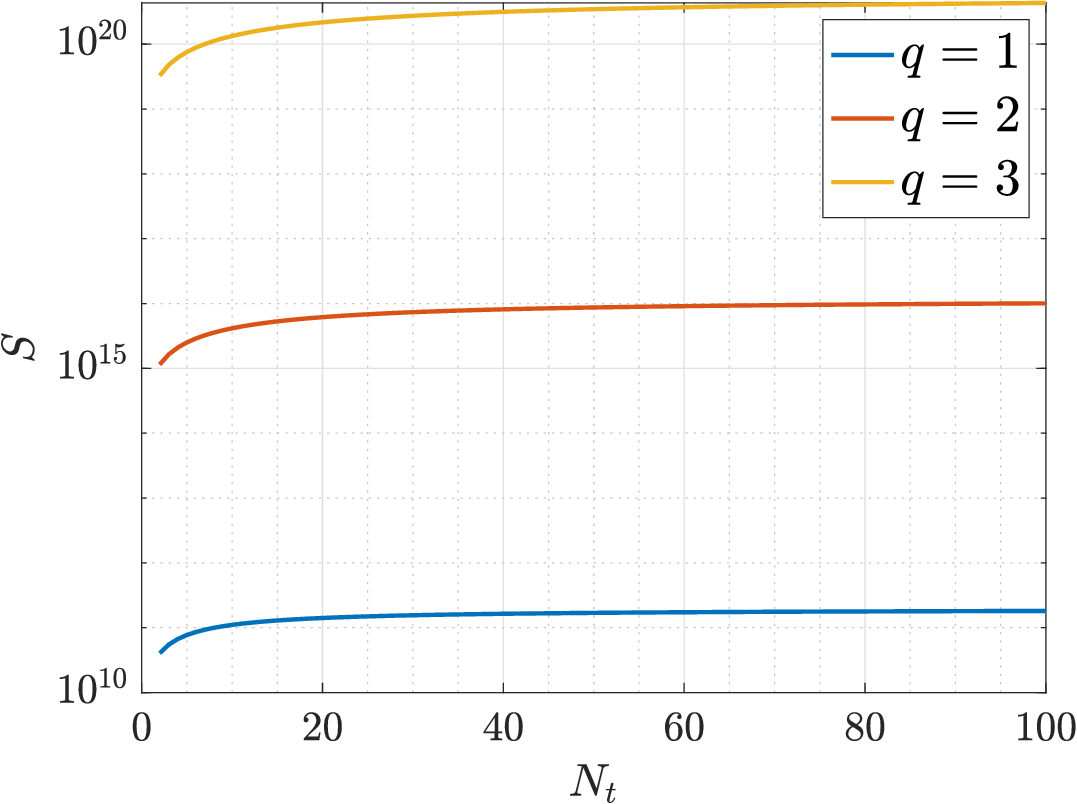}
\caption{\textcolor{black}{The evolution of the ratio $S$ in equation \eqref{eq:S-factor} for different values of $N_t$ and $q$. We set $G_{\tau} = G_{\alpha} = 1$ and $G_{\theta} = G_{\phi} = 180$.}}
\label{fig:S-factor}
\end{figure}

\textcolor{black}{In order to study the gains in complexity with respect to the \ac{MLE}, we plot $S$ for different values of number of targets $q$ in Fig. \ref{fig:S-factor} on a semi-log scale due to the large values involved.
For a fair assessment, we have specified $G_\alpha = G_\tau = 1$ which can reflect a case where $\alpha$ and $\tau$ \textit{are known} to the \ac{MLE}. Furthermore, we set a reasonable value of $G_\phi = G_\theta = 180$, which reflects a grid of step size of nearly $1^\circ$ if the search of both \ac{AoA} and \ac{AoD} is from $-90^\circ$ to $90^\circ$.
The exponential growth of values for various $N_t$ is evident in Fig. \ref{fig:S-factor}. Specifically, we observe magnitudes on the order of $10^{10}$ for a single target and $10^{15}$ for $q=2$ targets.

%Upper bounding $q$ by $N_r N_t$, which is the order of maximum number of resolvable sources in typical MIMO-radar systems, we get $S = \mathcal{O}(\zeta G_\tau^q G_\theta^q G_\phi^q G_\alpha^{2q})$, where $\zeta = \frac{N_P^4}{N_rN_t}$. It is evident that for any number of transmit and receive antennas, the complexity of \ac{MLE} with respect to the proposed parameterized one is dominated by the exponential increase in the number of targets on grid-search factors. For the sake of presentation, consider that all grids are of the same order $G$, therefore $S = \mathcal{O}(\zeta G^{5q})$, so even for a single target the complexity of \ac{MLE} is dominated by the size of $G$ which for a reasonable resolution can be of the order $10^2$, thus $S$ will be in the order of $10^{10}$. For two targets, this gain grows exponentially to $10^{20}$.
}

\section{Numerical Evaluation}
\label{sec:simulations}
In the scope of this work, we consider $N_{t} = 8$ transmit antennas and $N_{r} = 10$ receive antennas arranged in a uniform linear array (ULA) configuration with a half-wavelength spacing $\frac{\lambda}{2}$. 
%
% The carrier frequency is set to $f_c = \SI{30}{\giga\hertz}$. For\ac{OFDM} transmissions, we adopt a subcarrier spacing of $\Delta_{f} = \SI{960}{\kilo\hertz}$. \hl{These are unrealisitic numbers, please fix them or remove them from here if they are not relevant. You can alternatively write the bandwidth which is 61.4 MHz} %resulting in a symbol duration of $T = \frac{1}{\Delta_{f}} =  \SI{1.0417}{\micro\second}$. 
%Also, the cyclic prefix is set to $T_{cp} = \frac{T}{4} = \SI{0.2604}{\micro\second}$ bringing the total symbol duration to $T_{o} = T + T_{cp} = \SI{1.302}{\micro\second}$. 
A bandwidth of $\SI{61.44}{\mega\hertz}$ is used. The number of active subcarriers per symbol is $N _P = 64$ and the number of \ac{OFDM} symbols transmitted by the BS is $K_P = 10$, assumed to be known at the radar unit.
Both the BS and the radar unit are situated in fixed positions, while the targets are randomly positioned at varying distances, resulting in different \ac{AoA} and \ac{AoD} values. To capture the variability of the channel, we conduct experiments in a Monte Carlo fashion where each trial generates an independent realization of the channel.
\textcolor{black}{To allow duplicability of our work, 
the code has been made publicly available and is accessible on GitHub via the following link
\href{https://github.com/salmane-s9/Bistatic_ISAC}{https://github.com/salmane-s9/Bistatic\_ISAC}.

}
\subsection{\Ac{SNR} Analysis}
\label{sec:snr-analysis}
\begin{figure}[!t]
\centering
\includegraphics[width=3.5in]{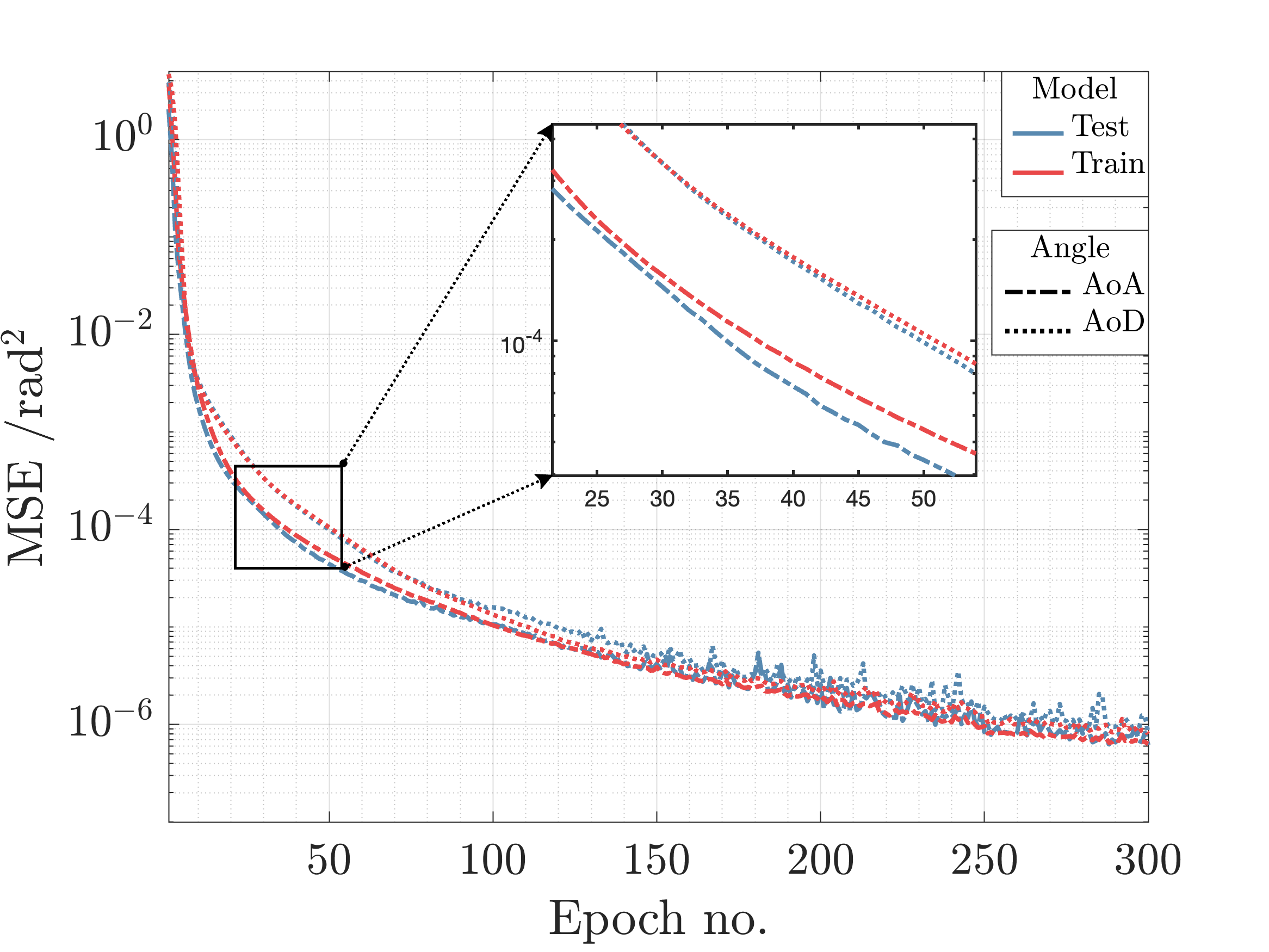}
\caption{Illustration of training and validation curves for AoA and AoD.}
\label{fig_train_test}
\end{figure}

In this section, we analyze the performance of the feed-forward network trained with different \ac{SNR} values. To achieve this, we systematically train the \ac{MLP} architecture using simulation data generated at specific \acp{SNR}. We then evaluate the network's performance on a test dataset with a range of \ac{SNR} values from $-5$ to $30 \dB$. Our goal is to identify the optimal training strategy that maximizes the network's performance across various scenarios. For training all complex models, we employed the backpropagation algorithm along with stochastic gradient descent and Adam optimizer. 
The training process consists of 300 epochs, during which we assess our model's performance and investigate overfitting by analyzing the training and validation \ac{MSE} losses. Fig.~\ref{fig_train_test} demonstrates that both loss curves steadily decrease over the course of training, confirming effective learning and generalization to unseen data. To address overfitting concerns, we adjust the learning rate schedule, starting at $1e^{-4}$ and decreasing it by a factor of 2 at epochs 200 and 250.
\begin{figure}[!t]
\centering
\includegraphics[width=3.5in]{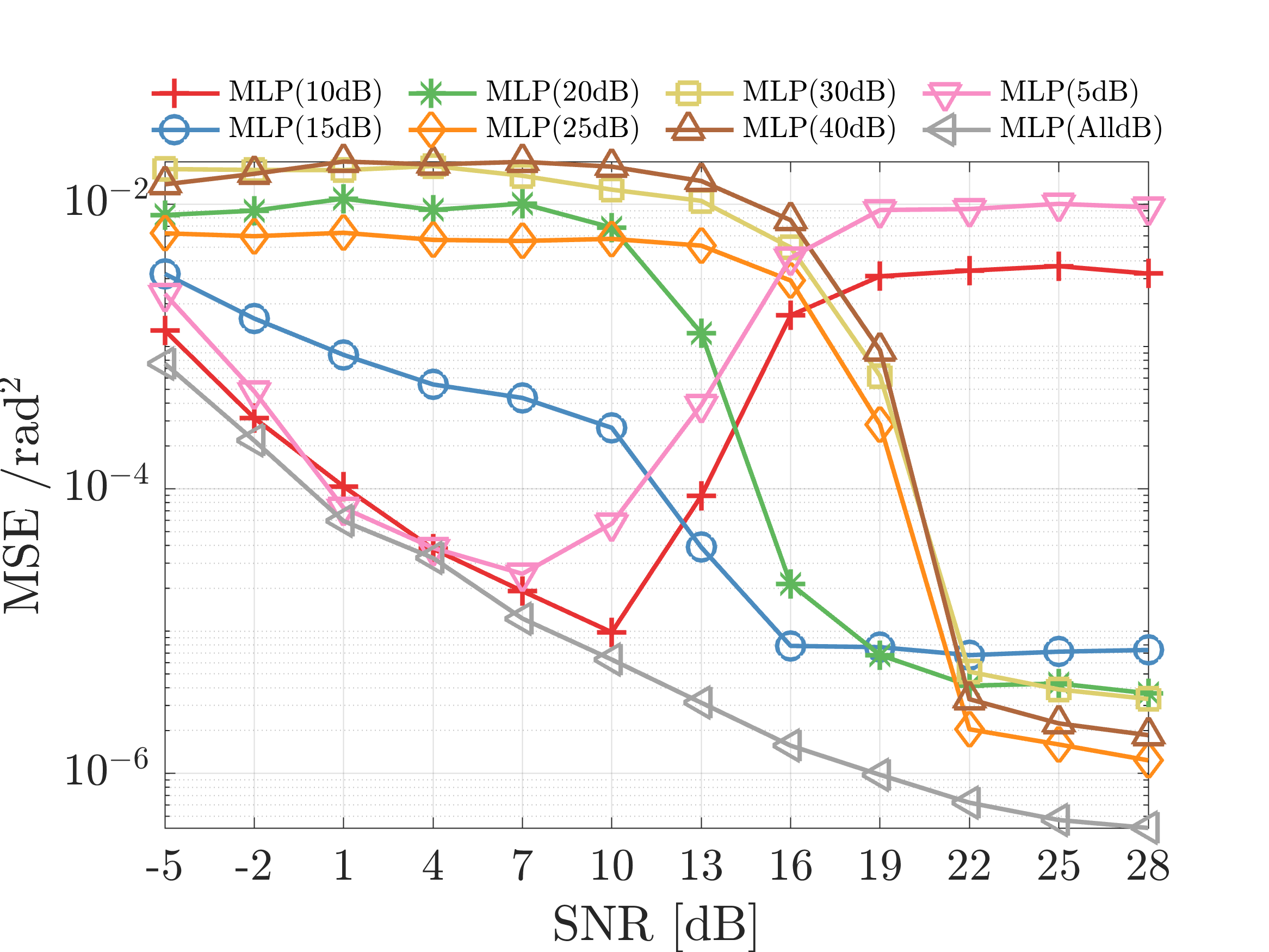}
\caption{Performance of the \ac{MLP} architecture trained at different \ac{SNR} values.}
\label{fig_snr_analysis}
\end{figure}
Fig.~\ref{fig_snr_analysis} presents the investigation of the trade-off between training the DL network on a wide range of \acp{SNR} and focusing on specific \ac{SNR} values. The results clearly demonstrate the significant impact of the \ac{SNR} used during training on the network's performance. Notably, the \ac{MSE} is lower when evaluating the network on simulations that closely match the \ac{SNR} it was trained on. Furthermore, our findings emphasize the effectiveness of training the network on a combination of simulation data with different \ac{SNR} values. This approach enhances the network's robustness, resulting in improved performance across all \ac{SNR} levels, as it is the case for the \textbf{MLP\_AlldB} algorithm that was trained on a dataset with the following \ac{SNR} values of $\left[5, 10, 15, 20, 25, 30, 40\right]$ dB. This can be explained by the fact that the model has effectively learned to adapt and perform well in various levels of noise associated with different \ac{SNR} values. Based on these findings, we have decided to adopt a comprehensive training strategy in all our experiments. This strategy involves training the network using a combination of \ac{SNR} values.
\textcolor{black}{It is worth noting a non-monotonic trend in the \ac{MSE} results. 
Specifically, with a low training \ac{SNR}, for instance $5\dB$, the \ac{MSE} initially decreases, as expected, reaching a minimum at about $5\dB$, and then gradually begins to rise, eventually leveling off at approximately $10^{-2} \rad^2$ MSE.
This phenomenon can be elucidated by overfitting, occurring particularly at low \ac{SNR}. In such cases, the model excessively adapts to the noise, aiming to minimize \ac{MSE} around the training \ac{SNR}.
On the other hand, when we increase the training \ac{SNR} towards an acceptable range, i.e. beyond $20\dB$, the overall \ac{MSE} performance exhibits a more regular "waterfall" structure.
Training at lower \acp{SNR} allows for a lower waterfall sensing threshold yet at the expense of a slightly lower MSE performance at higher \acp{SNR}.
Interestingly, when training is conducted under a diversified set of training \acp{SNR}, i.e. $5, 10 \ldots 40 \dB$, the best \ac{MSE} performance can be attained. }
\subsection{Performance Analysis}
\label{sec:performance-analysis}
\begin{figure}[!t]
\centering
\includegraphics[width=3.5in]{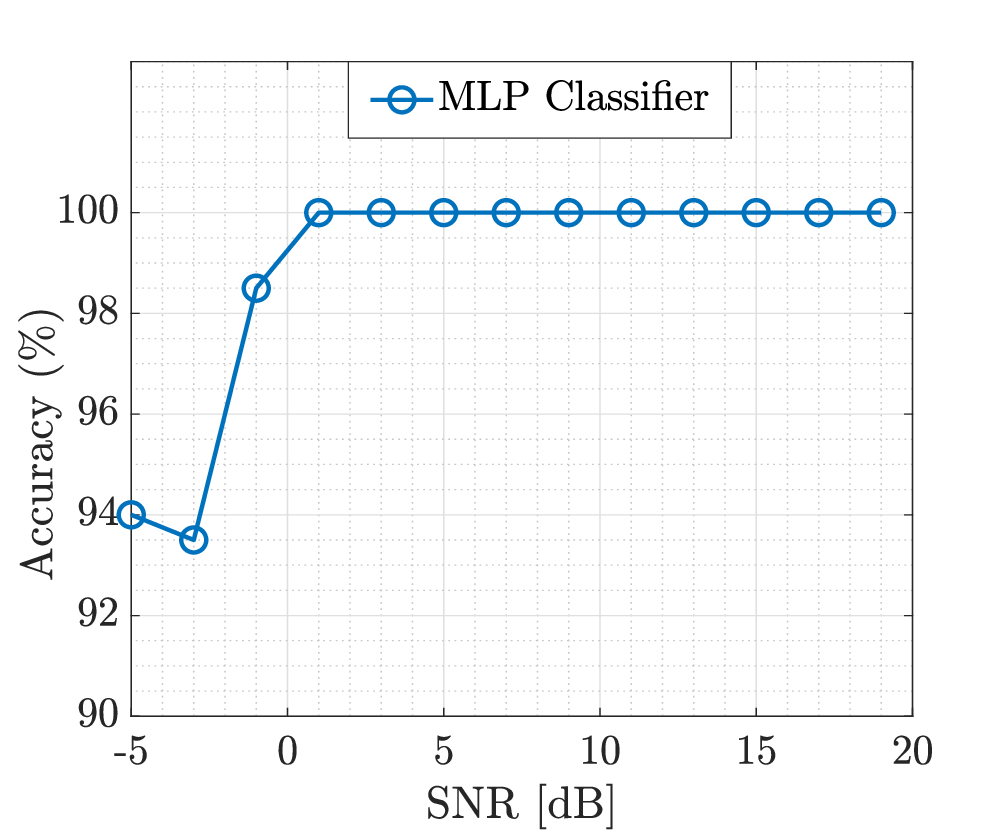}
\caption{Classification accuracy of the \ac{MLP} model for predicting the number of targets present within a peak.
\textcolor{black}{The \ac{MLP} classifier was trained on the following \ac{SNR} values $\left[-10, -5, 5, 15 \right]$ and tested on scenarios with \acp{SNR} ranging from $-5$ to $20$ \si{\dB}.}}
\label{fig_classification_accuracy}
\end{figure}
This section presents the performance behavior of the proposed \ac{NN} architecture in comparison to the benchmark $2$D estimation algorithm presented in Section~\ref{sec:benchmarking_algorithm}. As discussed in Section~\ref{subsec:coarse-timing-estimation}, there exists a scenario where multiple paths associated with different targets may arrive within the same discrete-time delay. Consequently, the $i_k$-th row of the matrix $\pmb{F}^H \pmb{\bar{{H}}}^T$, corresponding to the peak, is containing information about the specific targets' \ac{AoA} ($\theta_k$) and \ac{AoD} ($\phi_k$). Given that \ac{NN} architectures require prior knowledge of targets to predict their directions of arrival, one of the main objectives of this analysis is to assess the \ac{MLP} architecture's capability to predict the number of targets within a peak. First, we evaluate the classification accuracy of the proposed \ac{NN} by changing its output layer to softmax probabilities corresponding to the number of targets present within discrete-time delay. The results reported in Fig.~\ref{fig_classification_accuracy} show that this approach can effectively help classify the presence of one or multiple targets based on the peaks in the \ac{IFFT}, where the \ac{NN} was trained on simulation data with different \ac{SNR} values. 
Indeed, the classification accuracy reaches approximately 100\% for \ac{SNR} values greater than 0 dB. It is worth noting that we consider scenarios where at most five targets can be present in peak and therefore fixed the output size of the softmax layer to $N_{out} = 5$. 
\begin{figure}[!t]
\centering
\includegraphics[width=3.5in]{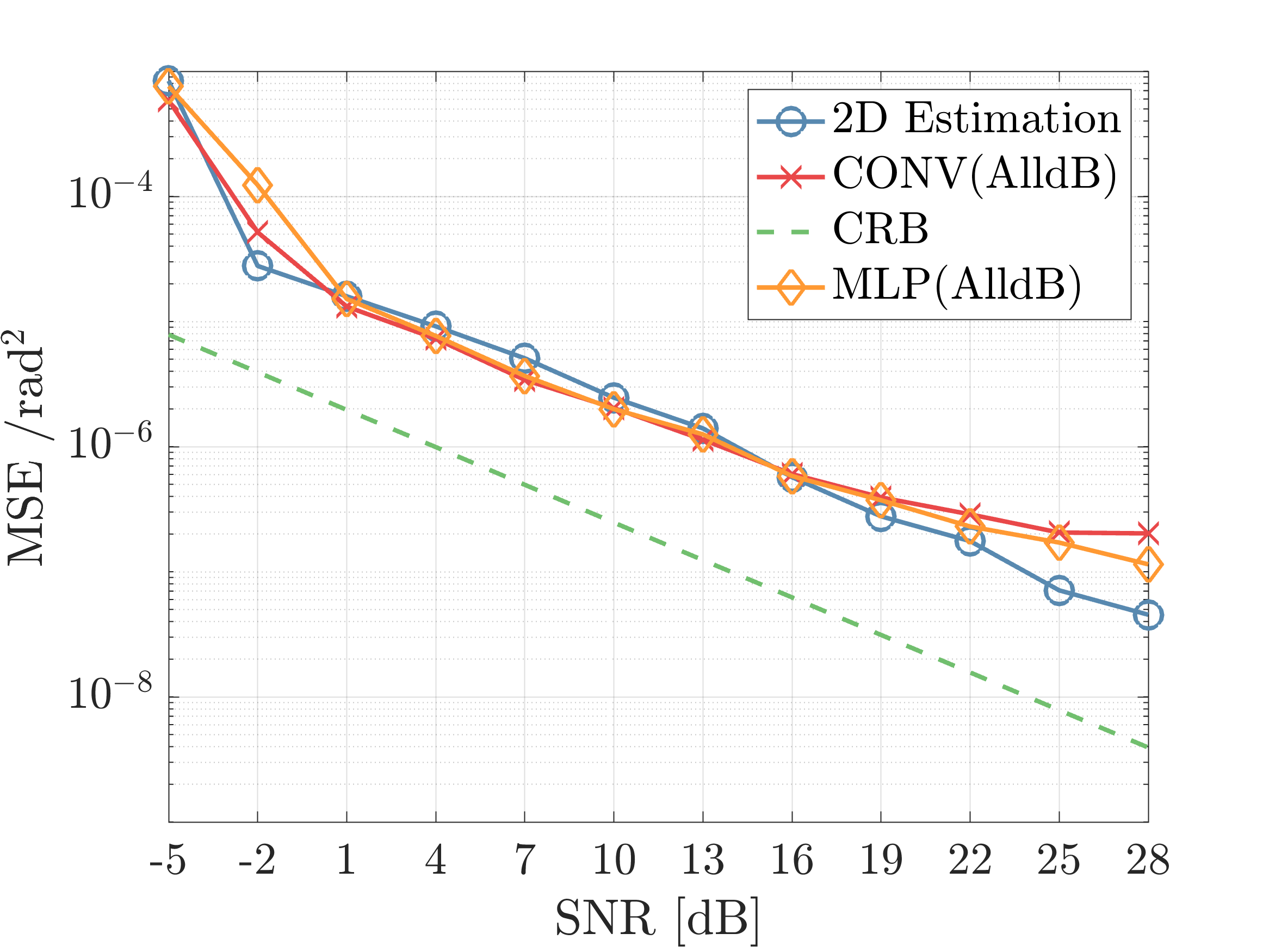}
\caption{Performance comparison of \ac{MLP} and convolutional networks with the parametric $2$D Estimation algorithm for \ac{AoA} estimation.}
\label{fig_1TarPerPeak}
\end{figure}
\begin{figure}[!t]
\centering
\includegraphics[width=3.5in]{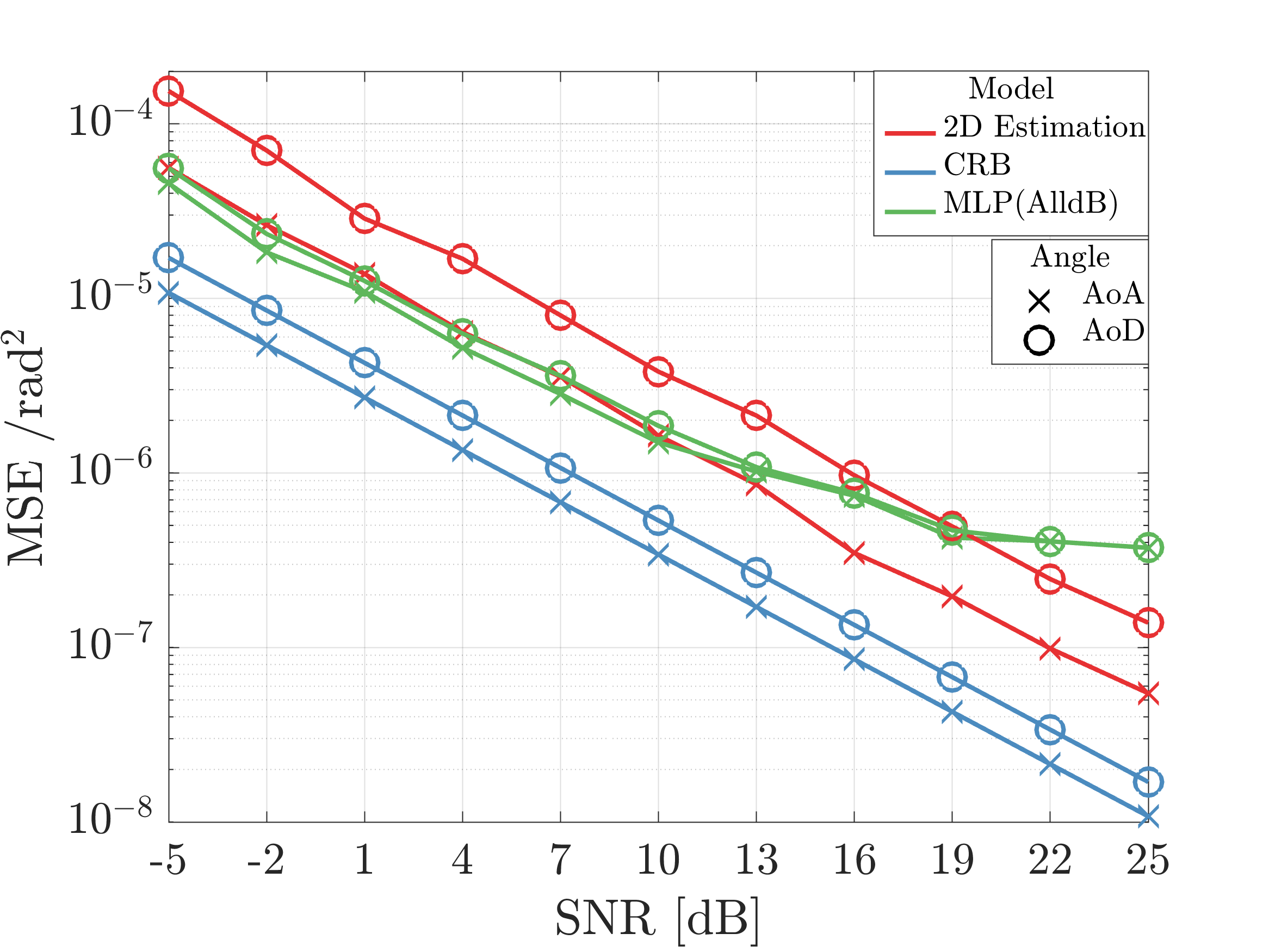}
\caption{Performance comparison of \ac{MLP} with the $2$D Estimation algorithm for both \ac{AoA} and \ac{AoD} angles for settings with two targets per peak.}
\label{fig_2TarPerPeak}
\end{figure}
Next, we compare the performance of both the convolutional and feed-forward \acp{NN} against the $2$D Estimation algorithm to evaluate their ability to accurately estimate \ac{AoA} and \ac{AoD} in two distinct scenarios. 
The first case shown in Fig.~\ref{fig_1TarPerPeak} considers a scenario where only one target is present within a peak. The second scenario, illustrated in Fig.~\ref{fig_2TarPerPeak}, involves two targets within the same peak. The results in both cases exhibit a good performance of the \ac{MLP} algorithm compared to the $2$D Estimation model, indicating its effectiveness for sensing estimation under various \ac{SNR} regimes. However, it is worth noting that there is a noticeable decrease in performance for high \ac{SNR} values, necessitating further investigation and potential optimization to enhance its performance in such scenarios.
The \ac{CRB} bound is given in Appendix \ref{app:appendix-CRB}. We also observe that for a requirement of $10^{-6}$ \ac{MSE} per radian$^2$, all methods are at about $9\dB$ \ac{SNR} away from the \ac{CRB} bound of the \ac{AoA}s. 
This gap is due to the complexity-performance trade-off with respect to the optimal \ac{MLE} estimator. 
The sub-optimality can be also seen from the timing criterion, whereby coarse estimates are being produced. Such estimates can, in turn, impact the performance at the price off reduced complexity. 
\textcolor{black}{Moreover, notice that the \ac{MLP} approach underperforms the parametric estimation one when $\SNR$ goes beyond $13\dB$.
The explanation posited is that when the noise level is very low, neural networks may not find it advantageous to learn more complex features, as there is already sufficient signal in the data. 
It is worth noting that neural networks are demonstrated to exhibit smooth functions in such scenarios.
The observed saturation in the \ac{DL} technique can be explained by its preference of fitting simple functions when there is enough signal. 
However, pushing training too far, i.e. very high number of epochs, may improve \ac{DL} performance. 
In addition, \textit{very deep networks} can be utilized to overcome saturation but at the cost of increased computational complexity \cite{pmlr-v97-rahaman19a}.}
\subsection{Complexity Analysis}
\begin{figure}[!t]
\centering
\includegraphics[width=3.5in]{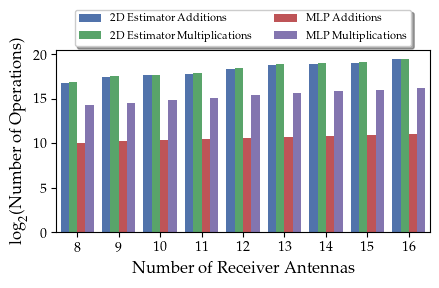}
\caption{Complexity comparison in terms of the number of multiplications and additions for both \ac{MLP} and $2$D estimation algorithm.}
\label{fig_complexity_Analysis}
\end{figure}

In this section, we provide a comprehensive analysis of the computational complexity associated with joint \ac{AoA} and \ac{AoD} estimation using the \ac{MLP} and parameterized benchmark $2$D algorithm. 
Our focus is primarily on quantifying the number of additions and multiplications required for these estimation methods. 
Additionally, we highlight that our analysis does not include the complexity associated with the channel estimation and coarse timing estimation steps, as they are shared by both algorithms and are thus excluded from the analysis.
Also, it is important to note that in the case of an environment with $q$ well-resolved targets, the computational complexity reported in \eqref{equ:MLPComplexitymul} and \eqref{equ:MLPComplexityadd} is multiplied by a factor of $q$ due to the increased batch size of the input, which corresponds to the number of targets present in the scene. 
In Fig.~\ref{fig_complexity_Analysis}, we provide the findings of our complexity study, which show the overall number of operations required as a function of the number of receive antennas. The results are obtained for an environment with 2 targets and a fixed number of transmit antennas $N_t = 8$. 
As the number of receive antennas increases, we observe a slower rate of increase in the total number of operations for the \ac{MLP} architecture compared to the parametric $2$D estimation algorithm. Indeed, the $2$D algorithm requires 6.5 and 10.3 times more multiplications than the \ac{MLP} for 8 and 16 receive antennas, respectively. Therefore, the latter only requires 15.27\% and 9.6\% of the computational complexity required by the $2$D algorithm for multiplications. This discrepancy in the rate of increase underscores the potential advantages of the \ac{MLP} architecture in terms of computational efficiency and scalability, making it practical for resource-constrained scenarios. However, it is important to note that these advantages come at a trade-off with a slight degradation in performance, particularly at higher \ac{SNR} values.

\begin{figure*}[!t]
\centering
\subfloat[\textcolor{black}{Normalized beampattern ($\Gamma_t = 1$) for different values of $\beta_t$}]{\includegraphics[width=3in]{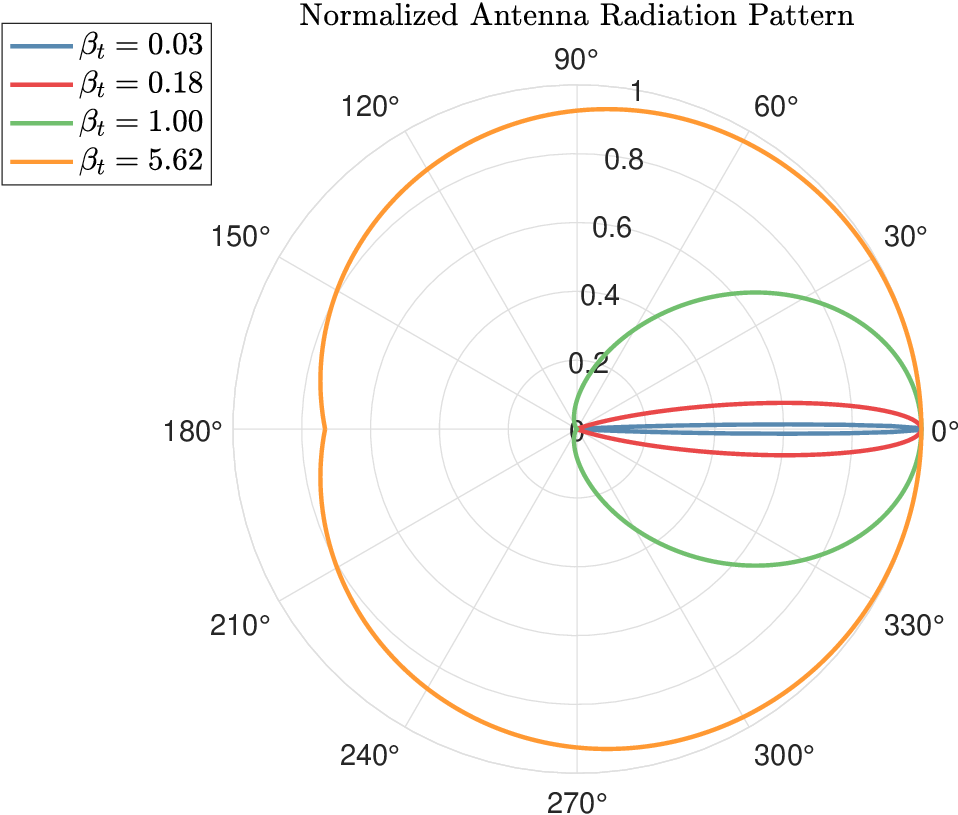}
\label{fig:ARP-beta}}
\hfil
% Blueprints/job00004_7
\subfloat[\textcolor{black}{\ac{CRB} as function of $\beta_t$ for different $\SNR$ values}]{\includegraphics[width=3in]{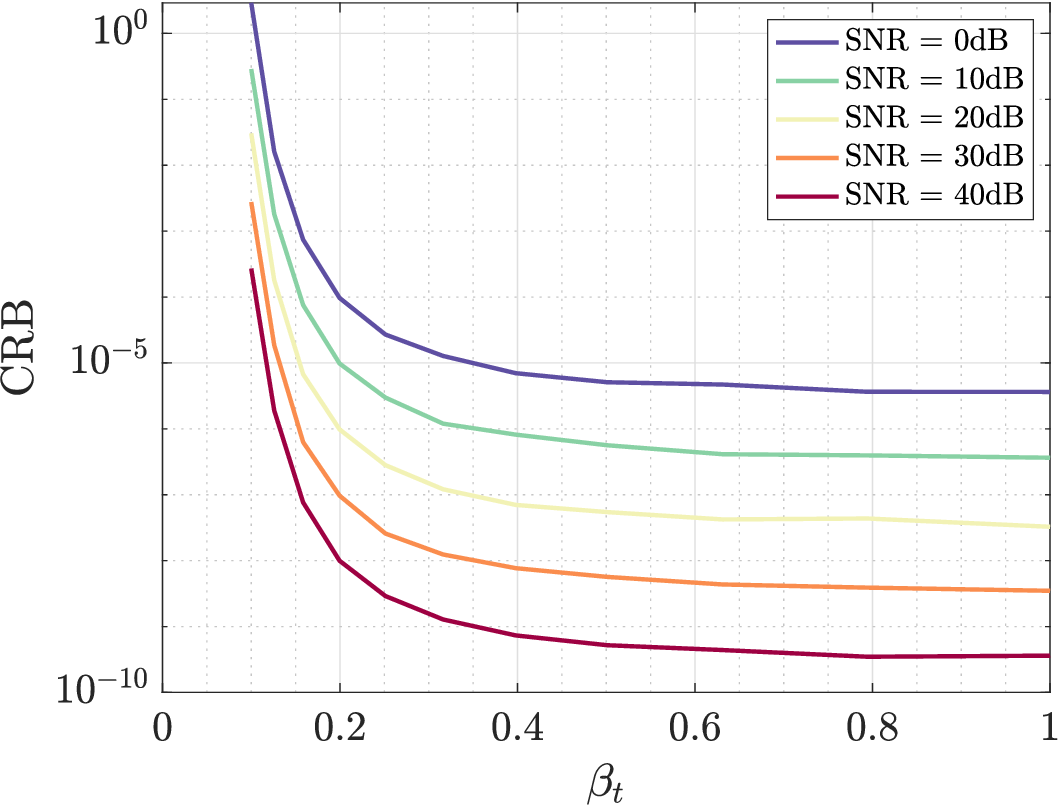} 
\label{fig:CRB-beta}}
\caption{\textcolor{black}{Antenna radiation pattern effect on the \ac{CRB} for $q = 1$ at $\theta_1 = 0^\circ$ and $\phi_1 = -15^\circ$.}}
\label{fig:ARP-CRB}
\end{figure*}

\subsection{\textcolor{black}{Antenna Radiation Pattern Impact}}
\label{sec:ARP-CRB}

\textcolor{black}{In Fig. \ref{fig:ARP-CRB}, we study the impact of antenna radiation pattern on system performance, in particular the \ac{CRB}.
First, we use a well-known approximating antenna radiation pattern, where the main beam uses a Gaussian-like shape as \cite{5336796}
\begin{equation}
	g_t(\phi)=\Gamma_t \exp \left(\frac{-[\mathcal{M}(\phi -\phi_{t,0})]^2}{\beta_t^2}\right),
\end{equation}
where $\Gamma_t$ is scaled to reflect the transmitted power and 
$\mathcal{M}(\phi)=\bmod _{2 \pi}(\phi+\pi)-\pi$ is restricted to live in $[-\pi,\pi]$.
In addition, the main beam points towards $\phi_{t,0}$ and $\beta_t$ specifies the beamwidth.
Note that index $t$ emphasizes that this antenna radiation pattern is dedicated for the \textit{"transmitter"}.
Hence, a similar definition could be used to describe the antenna radiation pattern at the receiver, i.e. the radar unit.
This approximation is very accurate for real practical antennas, such as those measured by \acp{LWA} \cite{7123663}.
In fact, \acp{LWA} exhibit several distinctive advantages,  including a relatively high gain, ease of fabrication, a broad bandwidth, and an inherent capability for beam scanning facilitated by a simple feed network. These attributes position LWAs as promising candidates for utilization in \ac{mmWave} applications \cite{9664489}.
For instance, the work in \cite{8292713} designed and simulated a \ac{mmWave} beam-steering \ac{LWA} hexagonal patch intended for the band $24$-$30$ GHz.
In Fig. \ref{fig:ARP-beta}, the aim is to show the resulting normalized beampattern as a function of $\beta_t$.
It is clear that increasing $\beta_t$ contributes to an inflation in the main beam of the antenna radiation pattern. For instance, as soon as $\beta_t$ exceeds $5.6$, the beampattern approaches an omni-directional structure.
In Fig. \ref{fig:CRB-beta}, we study the impact of $\beta_t$ on the \ac{CRB} for a single target, i.e. $q = 1$ located at $\theta_1 = 0^\circ$ and $\phi_1 = -15^\circ$.
A consistent trend of the \ac{CRB} can be reported for any $\SNR$ when $\beta_t$ is increased.
For instance, setting a \ac{CRB} target performance of $10^{-5} \rad^2$, the required beamwidth is 
$\beta_t = 0.35$ for $\SNR = 0 \dB$, whereas it is
$\beta_t = 0.2$ for $\SNR = 10 \dB$, and 
$\beta_t = 0.15$ for $\SNR = 20 \dB$.
This suggests that a more pointy beam requires higher \ac{SNR} to achieve a given \ac{CRB} for target sensing applications.
Indeed, one can improve the performance of pointy beams by integrating a beamsteering solution, i.e. by rotating $\phi_{t,0}$ towards sectors where targets may fall in.}

%
% \hl{Question from Marwa: can we say the complexity of the proposed MLP algorithm represents only 14.29\% of the $2$D algorithm when 8 receive antennas are used, and as low as 2.56\% for 16 antennas.} 
% {\color{blue}Answer from Roberto: I am not sure, I am a bit confused with the meaning of 700\% and 3900\% decrease. I cant read this from the graph. Is it multiplications? If so, say it in the text. Also, is it like the $2$D estimation has respectively 7 and 39 times larger complexity? If yes, then the numbers would be 14.29\% and 2.56\% respectively.

% My recommendation, say something like: the $2$D algorithm requires X and Y times more multiplications than the MLP for 8 and 16 receive antennas, respectively. This means that the MLP algrithm requires only 100/X\% and 100/Y\% of the complexity required by the $2$D algorithm.} 
%
\section{\textcolor{black}{Conclusion and Future Work}}\label{sec:conclusion}
In this paper, two methods for joint \ac{AoA} and \ac{AoD} estimation for bistatic \ac{ISAC} systems are presented. 
One solution is a \ac{DL}-based approach which leverages a complex-valued \ac{NN} and incorporates a preprocessing step involving coarse timing estimation, resulting in a reduced input size and improved computational efficiency. 
The second method is a parameterized solution that takes knowledge of the model, which is used as a benchmark.
The DL-based approach demonstrates competitive performance when compared to the parameterized method, while requiring a lower online computational complexity in terms of the number of additions and multiplications. 
Moreover, the \ac{DL} algorithm is able to accurately estimate the number of targets whose delays arrive at the same discrete-time delay, making it suitable for dynamic environments.
%we achieve perfect accuracy in estimating the number of targets within IFFT peaks when delays are not resolvable. 
%Overall, our comparison analysis highlights the potential practical application of our algorithm for sensing estimation in bistatic ISAC systems.
Nonetheless, while the performance gap with the benchmark method is generally acceptable, further research and optimization of the complex \ac{NN} are necessary to enhance performance, particularly in scenarios with high \ac{SNR}. %Other potential research directions include extensions towards more realistic system models that account for path loss and synchronization offsets between the base station and radar unit. Additionally, exploring the estimation of other parameters such as ToA and Doppler jointly with the directions of arrival holds promise for future research.

\textcolor{black}{It is worth noting that the existing implementations of both the \ac{DL}  technique and the parameterized-based method for 2D \ac{AoA} and \ac{AoD} estimation necessitate the re-execution of the methods whenever new target estimates are required.
Recognizing this limitation, our forthcoming research endeavors will focus on the development of tracking mechanisms within both the \ac{DL} and parametrized approaches. 
This deliberate integration is designed to simplify the process of updating \ac{AoA} and \ac{AoD} estimates, eliminating the requirement for sensing parameter re-estimation, especially in dynamic scenarios. }

%\small
\appendices
\label{sec:appendix}
\section{Cram\'er-Rao Bound Expressions}
\label{app:appendix-CRB}
The \ac{FIM} is given as follows
\begin{equation}
    \label{eq:FIM}
    \begin{split}
    \pmb{\Gamma}
   & \triangleq
    \mathbb{E} \Big[\frac{\partial \mathcal{L}(\pmb{\xi} )}{\partial \pmb{\xi} }  \frac{\partial\mathcal{L}(\pmb{\xi} )}{\partial \pmb{\xi} }^T \Big],
    \end{split}
\end{equation}
where $\pmb{\xi}$ is the vector of unknown parameters, 
\textcolor{black}{namely $\pmb{\xi} = \begin{bmatrix}
    \sigma & \pmb{\Theta} & \pmb{\Phi} & \pmb{\tau} & \bar{\breve{\pmb{\alpha}}} & \tilde{\breve{\pmb{\alpha}}}
\end{bmatrix}$,
where $\bar{\breve{\pmb{\alpha}}}$ is the real-part of $\breve{\pmb{\alpha}}$ 
and $\tilde{\breve{\pmb{\alpha}}}$ is the imaginary-part of $\breve{\pmb{\alpha}}$.}
Furthermore, $\mathcal{L}(\pmb{\xi} )$ is the log-likelihood of the model, i.e. $\mathcal{L}(\pmb{\xi} ) = \log f(\pmb{\mathcal{Y}})$ and $ f(\pmb{\mathcal{Y}})$ is the \ac{PDF} of the observed data defined in \eqref{eq:likelihood}.
The \ac{FIM} is partitioned according to the unknown variables, i.e. for any two parameter quantities, $\pmb{\Gamma}_{\pmb{a},\pmb{b}} =  \mathbb{E} [\frac{\partial \mathcal{L}(\pmb{\xi} )}{\partial \pmb{a} }  \frac{\partial\mathcal{L}^T(\pmb{\xi} )}{\partial \pmb{b} } ]$. Note that it is easy to see $\pmb{\Gamma}_{\sigma,\sigma} = \frac{N_r NK}{\sigma^4} $ and $\pmb{\Gamma}_{\sigma,\pmb{\Theta}}
=
\pmb{\Gamma}_{\sigma,\pmb{\Phi}}
=
\pmb{\Gamma}_{\sigma,\pmb{\tau}}
=
\pmb{\Gamma}_{\sigma,\bar{\breve{\pmb{\alpha}}}}
=
\pmb{\Gamma}_{\sigma,\tilde{\breve{\pmb{\alpha}}}}
=
\pmb{0}^T$
% \begin{align}
% \pmb{\Gamma}_{\sigma,\sigma} &= \frac{N_r NK}{\sigma^4} \\
% \pmb{\Gamma}_{\sigma,\pmb{\Theta}}
% &=
% \pmb{\Gamma}_{\sigma,\pmb{\Phi}}
% =
% \pmb{\Gamma}_{\sigma,\pmb{\tau}}
% =
% \pmb{\Gamma}_{\sigma,\bar{\pmb{\alpha}}}
% =
% \pmb{\Gamma}_{\sigma,\tilde{\pmb{\alpha}}}
% =
% \pmb{0}^T
% \end{align}
Now, denoting $\pmb{\Xi}_i = \pmb{a}_r(\theta_i) \pmb{a}_t^T(\phi_i)$, $\pmb{\Xi}_i^r = \pmb{d}_r(\theta_i) \pmb{a}_t^T(\phi_i)$ and $\pmb{\Xi}_i^t = \pmb{a}_r(\theta_i) \pmb{d}_t^T(\phi_i) $, where $\pmb{d}_r(\theta) = \frac{\partial \pmb{a}_r(\theta)}{\partial \theta }$ and $\pmb{d}_t(\phi) = \frac{\partial \pmb{a}_t(\phi)}{\partial \phi }$ are the partial derivatives of the receive and transmit steering vectors with respect to $\theta$ and $\phi$, respectively.
\textcolor{black}{In addition, we define $\breve{\alpha}_i = \alpha_ig_t(\phi_i)g_r(\theta_i)$.}
To this end, we summarize the \ac{FIM} block-matrices appearing in \eqref{eq:FIM} as follows. Derivation details in what follows are omitted due to lack of space. First, we compute all second-order partial derivatives whenever $\pmb{\Theta}$ appears, i.e.
\textcolor{black}{\begin{align*}
 [\pmb{\Gamma}_{\pmb{\Theta},\pmb{\Theta}}]_{i,j}
    &=
    \frac{2}{\sigma^2}
    \sum\limits_{n,k}
    \Re
    \Big(
     \pmb{s}_{n,k}^H
    [\breve{\alpha}_i c_n(\tau_i) \pmb{\Xi}_i^r]^H
    [\breve{\alpha}_j c_n(\tau_j) \pmb{\Xi}_j^r]
    \pmb{s}_{n,k}
    \Big), \\ 
    [\pmb{\Gamma}_{\pmb{\Theta},\pmb{\Phi}}]_{i,j}
    &=
    \frac{2}{\sigma^2}
    \sum\limits_{n,k}
    \Re
    \Big(
    \pmb{s}_{n,k}^H
    [\breve{\alpha}_i c_n(\tau_i)  \pmb{\Xi}_i^r]^H
    [\breve{\alpha}_j c_n(\tau_j)  \pmb{\Xi}_j^t]
    \pmb{s}_{n,k}
    \Big), \\
    [\pmb{\Gamma}_{\pmb{\Theta},\pmb{\tau}}]_{i,j}
    &=
    \frac{2}{\sigma^2}
    \sum\limits_{n,k}
    \Re
    \Big(
    \pmb{s}_{n,k}^H
    [\breve{\alpha}_i c_n(\tau_i)  \pmb{\Xi}_i^r]^H
    [\breve{\alpha}_j d_n(\tau_j)  \pmb{\Xi}_j]
    \pmb{s}_{n,k}
    \Big), \\
    [\pmb{\Gamma}_{\pmb{\Theta},\bar{\breve{\pmb{\alpha}}}}]_{i,j}
    &=
    \frac{2}{\sigma^2}
    \sum\limits_{n,k}
    \Re
    \Big(
    \pmb{s}_{n,k}^H
    [\breve{\alpha}_i c_n(\tau_i)\pmb{\Xi}_i^r]^H
    [c_n(\tau_j) \pmb{\Xi}_j]
    \pmb{s}_{n,k}
    \Big), \\
    [\pmb{\Gamma}_{\pmb{\Theta},\tilde{\breve{\pmb{\alpha}}}}]_{i,j}
    &=
    \frac{2}{\sigma^2}
    \sum\limits_{n,k}
    \Re
    \Big(
    \pmb{s}_{n,k}^H
    [\breve{\alpha}_i c_n(\tau_i) \pmb{\Xi}_i^r]^H
    [j c_n(\tau_j) \pmb{\Xi}_j]
    \pmb{s}_{n,k}
    \Big),
\end{align*}}
where $d_n(\tau) = \frac{\partial c_n(\tau)}{\partial \tau}$. Then, we compute all second-order partial derivatives whenever $\pmb{\Phi}$ appears, i.e.
\textcolor{black}{\begin{align*}
    [\pmb{\Gamma}_{\pmb{\Phi},\pmb{\Phi}}]_{i,j}
    &=
    \frac{2}{\sigma^2}
    \sum\limits_{n,k}
    \Re
    \Big(
    \pmb{s}_{n,k}^H
    [\breve{\alpha}_i c_n(\tau_i) \pmb{\Xi}_i^t]^H
     [\breve{\alpha}_j c_n(\tau_j) \pmb{\Xi}_j^t]
    \pmb{s}_{n,k}
    \Big), \\ 
    [\pmb{\Gamma}_{\pmb{\Phi},\pmb{\tau}}]_{i,j}
    &=
    \frac{2}{\sigma^2}
    \sum\limits_{n,k}
    \Re
    \Big(
    \pmb{s}_{n,k}^H
    [\breve{\alpha}_i c_n(\tau_i) \pmb{\Xi}_i^t]^H
     [\breve{\alpha}_j d_n(\tau_j) \pmb{\Xi}_j]
    \pmb{s}_{n,k}
    \Big), \\ 
    [\pmb{\Gamma}_{\pmb{\Phi},\bar{\breve{\pmb{\alpha}}}}]_{i,j}
    &=
    \frac{2}{\sigma^2}
    \sum\limits_{n,k}
    \Re
    \Big(
    \pmb{s}_{n,k}^H
    [\breve{\alpha}_i c_n(\tau_i) \pmb{\Xi}_i^t]^H
     [ c_n(\tau_j) \pmb{\Xi}_j]
    \pmb{s}_{n,k}
    \Big), \\ 
    [\pmb{\Gamma}_{\pmb{\Phi},\tilde{\breve{\pmb{\alpha}}}}]_{i,j}
    &=
    \frac{2}{\sigma^2}
    \sum\limits_{n,k}
    \Re
    \Big(
    \pmb{s}_{n,k}^H
    [\breve{\alpha}_i c_n(\tau_i) \pmb{\Xi}_i^t]^H
     [ j c_n(\tau_j) \pmb{\Xi}_j]
    \pmb{s}_{n,k}
    \Big),
\end{align*}}
Following the above expressions, we compute all \ac{FIM} partial derivatives where $\pmb{\tau}$ appears
\textcolor{black}{\begin{align*}
    [\pmb{\Gamma}_{\pmb{\tau},\pmb{\tau}}]_{i,j}
    &=
    \frac{2}{\sigma^2}
    \sum\limits_{n,k}
    \Re
    \Big(
    \pmb{s}_{n,k}^H
    [\breve{\alpha}_i d_n(\tau_i) \pmb{\Xi}_i]^H
     [\breve{\alpha}_j d_n(\tau_j) \pmb{\Xi}_j]
    \pmb{s}_{n,k}
    \Big), \\ 
    [\pmb{\Gamma}_{\pmb{\tau},\bar{\breve{\pmb{\alpha}}}}]_{i,j}
    &=
    \frac{2}{\sigma^2}
    \sum\limits_{n,k}
    \Re
    \Big(
    \pmb{s}_{n,k}^H
    [\breve{\alpha}_i d_n(\tau_i) \pmb{\Xi}_i]^H
     [ c_n(\tau_j) \pmb{\Xi}_j]
    \pmb{s}_{n,k}
    \Big), \\ 
    [\pmb{\Gamma}_{\pmb{\tau},\tilde{\breve{\pmb{\alpha}}}}]_{i,j}
    &=
    \frac{2}{\sigma^2}
    \sum\limits_{n,k}
    \Re
    \Big(
    \pmb{s}_{n,k}^H
    [\breve{\alpha}_i d_n(\tau_i) \pmb{\Xi}_i]^H
     [ j c_n(\tau_j) \pmb{\Xi}_j]
    \pmb{s}_{n,k}
    \Big),
\end{align*}}
Next, we compute all partial derivatives where \textcolor{black}{$\bar{\breve{\pmb{\alpha}}}$} appears
\textcolor{black}{\begin{align*}
[\pmb{\Gamma}_{\bar{\breve{\pmb{\alpha}}},\bar{\breve{\pmb{\alpha}}}}]_{i,j}
    &
    =
    \frac{2}{\sigma^2}
    \sum\limits_{n,k}
    \Re
    \Big(
    \pmb{s}_{n,k}^H
    [c_n(\tau_i) \pmb{\Xi}_i]^H
     [c_n(\tau_j) \pmb{\Xi}_j]
    \pmb{s}_{n,k}
    \Big), \\
    [\pmb{\Gamma}_{\bar{\breve{\pmb{\alpha}}},\tilde{\breve{\pmb{\alpha}}}}]_{i,j}
    &
    =
    \frac{2}{\sigma^2}
    \sum\limits_{n,k}
    \Re
    \Big(
    \pmb{s}_{n,k}^H
    [c_n(\tau_i) \pmb{\Xi}_i]^H
     [j c_n(\tau_j) \pmb{\Xi}_j]
    \pmb{s}_{n,k}
    \Big),
\end{align*}}
Then we compute all partial derivatives where \textcolor{black}{$\tilde{\breve{\pmb{\alpha}}}$} appears
\textcolor{black}{\begin{equation}
    [\pmb{\Gamma}_{\tilde{\breve{\pmb{\alpha}}},\tilde{\breve{\pmb{\alpha}}}}]_{i,j} 
    =
    \frac{2}{\sigma^2}
    \sum\limits_{n,k}
    \Re
    \Big(
    \pmb{s}_{n,k}^H
    [j c_n(\tau_i) \pmb{\Xi}_i]^H
     [j c_n(\tau_j) \pmb{\Xi}_j]
    \pmb{s}_{n,k}
    \Big).
\end{equation}}
Now, the \ac{CRB} is obtained as follows
\begin{equation}
    \CRB (\pmb{\Theta}) 
    =
   [ \pmb{\Gamma}^{-1} ]_{2:(q+1),2:(q+1)}
\end{equation}
\begin{equation}
    \CRB (\pmb{\Phi}) 
    =
    [\pmb{\Gamma}^{-1}]_{(q+2):(2q+1),(q+2):(2q+1)}
\end{equation}

\bibliographystyle{ieeetr}
\bibliography{references_ha}{}

\vfill

\end{document}